\documentclass[aps,prb,10pt,twocolumn,superscriptaddress,longbibliography]{revtex4}

\usepackage{graphicx}
\usepackage{amssymb, amsmath}
\usepackage{color}
\usepackage{ulem}
\usepackage{bm}
\usepackage{graphicx}
\usepackage{subfigure}
\usepackage{dcolumn}
\usepackage{mathtools}
\usepackage{pifont}
\def\I{\uppercase\expandafter{\romannumeral 1}}
\def\II{\uppercase\expandafter{\romannumeral 2}}
\def\III{{\uppercase\expandafter{\romannumeral 3}}}
\def\IV{{\uppercase\expandafter{\romannumeral 4}}}
\def\V{{\uppercase\expandafter{\romannumeral 5}}}
\def\VI{{\uppercase\expandafter{\romannumeral 6}}}
\def\VII{{\uppercase\expandafter{\romannumeral 7}}}
\def\i{\lowercase\expandafter{\romannumeral 1}}
\def\ii{\lowercase\expandafter{\romannumeral 2}}
\def\iii{{\lowercase\expandafter{\romannumeral 3}}}
\def\iv{{\lowercase\expandafter{\romannumeral 4}}}
\def\v{{\lowercase\expandafter{\romannumeral 5}}}
\def\vi{{\lowercase\expandafter{\romannumeral 6}}}
\def\vii{{\lowercase\expandafter{\romannumeral 7}}}

\def\angstrom{\mbox{\normalfont\AA}}

\def\nn{\nonumber\\}

\def\angstrom{\mbox{\normalfont\AA}}

\def\nn{\nonumber\\}

\begin{document}

\title{Topological piezoelectric response in moir\'e graphene systems}

\author{Ran Peng}
\affiliation{School of Physical Science and Technology, ShanghaiTech University, Shanghai 200031, China}

\author{Jianpeng Liu}
\email[]{liujp@shanghaitech.edu.cn}
\affiliation{School of Physical Science and Technology, ShanghaiTech University, Shanghai 200031, China}
\affiliation{ShanghaiTech laboratory for topological physics, ShanghaiTech University, Shanghai 200031, China}

\begin{abstract}
We theoretically study the piezoelectric effects in moir\`e graphene systems. 
Since the strain couples to the electrons in the system  as a pseudo vector potential, which has opposite signs for the $K$ and $K'$ valleys of graphene, its effects on the two valleys with opposite Chern numbers do not cancel out, but adds up. As a result, some components of the piezoelectric tensor in these systems, which typically have non-trivial topology in their flat bands, are nearly quantized in terms of the valley Chern numbers. Such a conclusion is verified by numerical calculations of the in-plane piezoelectric response of hBN-aligned twisted bilayer graphene, twisted bilayer-monolayer graphene, and twisted double bilayer graphene systems using both continuum model and atomistic tight-binding model. We find that by tuning the vertical displacement field and/or twist angle, which may induce gap closures between the flat bands and remote bands in these systems,  plateau shapes of the  piezoelectric response are obtained,  with abrupt jumps across the topological phase transitions. We propose that such nearly quantized piezoelectric response may serve as a direct experimental probe for the valley Chern numbers of the flat bands in moir\'e graphene systems.       
\end{abstract}


\maketitle

The intriguing phenomena, such as superconductivity \cite{cao-nature18-supercond,dean-tbg-science19,marc-tbg-19, efetov-nature19,efetov-nature20,young-tbg-np20,li-tbg-science21,cao-tbg-nematic-science21}, correlated insulating states \cite{cao-nature18-mott,efetov-nature19,tbg-stm-pasupathy19,tbg-stm-andrei19,tbg-stm-yazdani19, tbg-stm-caltech19, young-tbg-science19,efetov-nature20,young-tbg-np20,li-tbg-science21}, and quantum anomalous Hall states \cite{young-tbg-science19, sharpe-science-19, efetov-arxiv20}, observed in twisted bilayer graphene (TBG) around magic angle, has drawn great attention in recent years. This attention soon extends to twisted multilayer graphene systems such as  twisted bilayer-monolayer graphene \cite{young-monobi-nature20,Yankowitz-monobi-np2020,shi-tbmg-np21} and twisted double bilayer graphene systems \cite{kim-tdbg-nature20,zhang-tdbg-np20,cao-tdbg-nature20,Pasupathy-nematic-tdbg-arxiv20}, and to hBN-graphene heterostructures \cite{chen-hbn-trilayer-nature19,chen-trilayer-hbn-mott-np19},  which exhibit equally interesting properties. All these exotic effects are believed to be attributed by the low-energy flat bands in these moir\'e graphene systems, which are found to be topologically non-trivial \cite{po-tbg-prb19, jpliu-prb19,zaletel-tbg-2019, song-tbg-prl19, yang-tbg-prx19, senthil-tbg-prr19,jpliu-prx19}. 

In magic-angle TBG, the low-energy states can be divided into two sets: one from the $K$ valley, and the other from the $K'$ valley of graphene, each of which covers the entire moir\'e Brillouin zone \cite{macdonald-pnas11,castro-neto-prb12}. Moreover, by virtue of the large moir\'e superlattice cosntant $L_s\sim10\,$nm, the intervalley scattering induced by the long-period moir\'e potential between the two sets of low-energy states is negligible \cite{macdonald-pnas11,castro-neto-prb12}, thus the total charge of each valley is approximately separately conserved characterized by valley $U(1)$ symmetry \cite{po-prx18}. This means that one can separately define topological quantities, such as Chern numbers\cite{jpliu-prx19,senthil-tbg-prr19}, and Wilson loops\cite{song-tbg-prl19,jpliu-prb19}, for each valley.  Indeed, the nontrivial topological properties in the flat bands of magic-angle TBG are  manifested as odd-winding Wilson loops \cite{song-tbg-prl19,yang-tbg-prx19,jpliu-prb19}, and  the flat bands would acquire nonzero and valley-contrasting Chern numbers when $C_{2z}$ symmetry is broken due to alignment of the hexagonal boron nitride (hBN) substrate  \cite{jpliu-prb19,senthil-tbg-prr19,zaletel-tbg-2019}.  
However, direct evidence of the non-trivial VCNs of the moir\'e graphene systems from experiments are still absent. On one hand, the expected gapless helical topological edge states from the nonzero VCNs would be gapped out by the significantly enhanced intervalley scatterings at the edges due to the atomic-scale potentials from the sample boundaries.
On the other hand,  without interaction-driven spontaneous valley polarization, the bands contributed by different valleys always intertwine with their time-reversal (TR) counterpart, so that they cannot be separated by tuning the carrier density, and the topological effects  from the two valleys cancel each other due to the opposite VCNs. This is true for most measures of tuning the materials, with one possible exception---strain. In low energy bands, strain does not act on the two valleys in equal way, but oppositely. A valley dependent pseudo vector potential can be induced by strain, which is opposite in the two valleys associated by TR operation. As a result, strain effect on different valleys does not cancel each other, but adds up. 

The above reasoning applies to generic twisted multilayer graphene (TMG) and other moir\'e graphene heterostructure systems, which typically have topological flat bands with nonzero VCNs that are highly tunable by displacement fields and twist angle \cite{jpliu-prx19,senthil-tbg-prr19,koshino-tdbg-prb19}. 
Such nontrivial topological properties motivate us to study the strain effects in this whole series of systems. In the continuum model framework, we argue that the piezoelectric response of these twisted graphene  systems are (nearly) quantized, in the sense that they can be expressed as a constant multiplied by the valley Chern numbers of the system. Taking hBN-aligned TBG, twisted bilayer-monolayer grpahene (TBMG), and twisted double-bilayer graphene (TDBG) as examples, we have numerically calculated the piezoelectric response based on both continuum model and atomic tight-binding models, and nearly quantized plateaus of piezoelectric tensors are obtained in all of the three systems.

\begin{figure}[!htbp]
\includegraphics[width=0.45\textwidth]{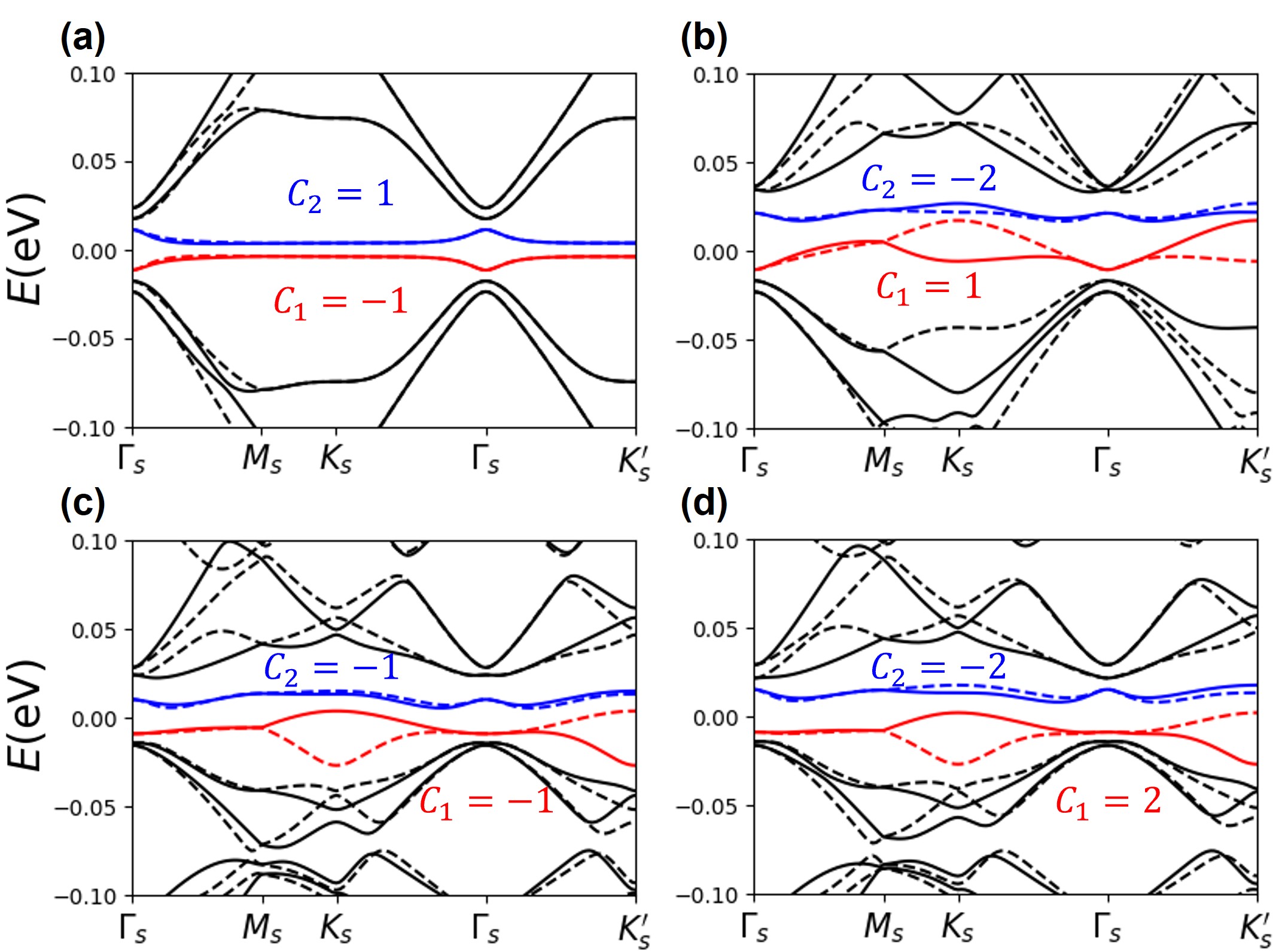}
\caption{~\label{fig1} Band structures of: (a) TBG with staggered sublattice potential $\Delta=15\,{\rm meV}$ at $\theta=1.05^\circ$, (b) twisted bilayer-monolayer graphene with vertical electrostatic potential energy difference  $U_d=-0.0536\,{\rm eV}$ (see text) at $\theta=1.2^\circ$, (c) AB-BA stacked TDBG with $U_d=-0.0402{\rm eV}$ at $\theta=1.2^\circ$, and (d) AB-AB stacked TDBG with $U_d=-0.0402{\rm eV}$ at $\theta=1.2^\circ$.} 
\end{figure}

\paragraph{Theories of polarization and piezoelectric response \textemdash}
According to the modern theory of polarization \cite{kingsmith-prb93, resta07}, polarization in an insulating crystal is measured by the charge current in an adiabatic evolution that establishes the final polarized state. The theory suggests a Berry-phase-like expression for the dependence of the polarization vector $\mathbf{P}$ on the adiabatic parameter $\lambda$: 
\begin{equation}
    \frac{\partial \mathbf{P}}{\partial \lambda} = \frac{e}{(2\pi)^2}\int {\rm d^2\mathbf{ k}} \sum_{n}\Omega_{\mathbf{k}\lambda,n},
\label{eq.Pdef}
\end{equation}
where
\begin{equation}
    \Omega_{\mathbf{k}\lambda,n}=i(\langle \nabla_{\mathbf{k}} u_{n{\mathbf{k}}} \vert \partial_{\lambda} u_{n{\mathbf{k}}} \rangle - \langle \partial_{\lambda} u_{n{\mathbf{k}}} \vert \nabla_{\mathbf{k}} u_{n{\mathbf{k}}} \rangle),
    \label{eq:berry-phase}
\end{equation}
and $\vert u_{n{\mathbf{k}}} \rangle$ refers to the periodic part of the $n^{\rm th}$ occupied Bloch function of the system. Note that Eq~(\ref{eq:berry-phase}) is the correct even for a band with nonzero Chern number, although there is subtlety in considering the filling of edge states of a Chern insulaotr in the adiabatic process \cite{polarization-chern}.

On the other hand, the Chern number of an isolated band $\vert u_{n{\mathbf{k}}} \rangle$ of a 2D crystal is expressed as 
\begin{equation}
    C_{n}=\frac{1}{2\pi i}\int {\rm d^2\mathbf{ k}} (\langle \partial_{k_x} u_{n{\mathbf{k}}} \vert \partial_{k_y} u_{n{\mathbf{k}}} \rangle - \langle \partial_{k_y} u_{n{\mathbf{k}}} \vert \partial_{k_x} u_{n{\mathbf{k}}} \rangle)
    \label{eq.Chern}
\end{equation}
The similarity between  Eq.~(\ref{eq.Chern}) and Eq.~(\ref{eq.Pdef}) indicates a possibility of topological effects to show up in 2D crystal polarization. Since strain couples to electrons as a pseudo vector potential, and the coupling coefficients have opposite signs for the opposite valleys ($K$ and $K'$),  we conjecture that the strain induced polarization, i.e., the piezoelectric response, may be the correct quantity to manifest nonzero VCNs of the moir\'e grahene systems. 

The piezoelectric response is normally known as the change of polarization $\mathbf{P}$ induced by the strain tensor $\mu_{jk}$: $\gamma_{ijk}=\partial P_i/\partial\mu_{jk}$, which turns out to be ``improper" due to possible ambiguities \cite{vanderbilt-piezo-00}. A proper definition of piezoelectric response tensor is  expressed as
\begin{equation}
    \gamma_{ijk}=\frac{\partial \dot{P}_i}{\partial \dot{\mu}_{jk}},
\end{equation}
where $\dot{P}_i$ and $\dot{\mu}_{jk}$ denote the time derivatives of  the polarization and  the strain, with $i,j,k=x,y$ for 2D systems. The strain tensor is defined by the shift $\Delta {\mathbf{r}}$ of a real-space position $\mathbf{r}$ with respect to the original position ${\mathbf{r}}$: $\Delta r_i=\sum_{j,k=x,y}\mu_{ij}r_j$.
By simply replacing the adiabatic parameter $\lambda$ in Eq.~(\ref{eq.Pdef}) with strain component $\mu_{jk}$,  and multiplying both sides of Eq.~(\ref{eq.Pdef}) by $\dot{\mu}_{jk}$,  we obtain the expression for the PET component contributed from the valley $\eta$ ($\eta=\pm$ refers to the $K$ and $K$' valleys) in the twisted graphene system:
\begin{equation}
    \gamma_{ijk}^{\eta} = \frac{e}{(2\pi)^2}\int {\rm d^2\mathbf{ k}} \,\Omega_{ijk}^{\eta},
\end{equation}\;
where
\begin{equation}
    \Omega_{ijk}^{\eta}=\sum_{n}\,i(\langle \partial_{k_i} u_{\eta,n{\mathbf{k}}} \vert \partial_{\mu_{jk}} u_{\eta,n{\mathbf{k}}} \rangle - \langle \partial_{\mu_{jk}} u_{\eta,n{\mathbf{k}}} \vert \partial_{k_i} u_{\eta,n{\mathbf{k}}} \rangle).
\label{eq.Omegaijk}
\end{equation}
Here the index $n$ refers to the occupied band of the $\eta$ valley. A small homogeneous strain applied to the system is equivalent to an effective vector poential $A_i^{\eta}=\xi_{ijk}^{\eta}\mu_{jk}$ in the $\eta$ valley \cite{bi-tbg-strain}, which is linearly coupled to the wavevector: ${\mathbf{k}}\rightarrow {\mathbf{k}}+{\mathbf{A}}^{\eta}$, and the partial derivative with respect to strain in Eq.~(\ref{eq.Omegaijk}) can be substituted by a partial derivative with respect to wavevector: $\partial_{\mu_{jk}}=\xi^{\eta}_{ijk} \partial_{k_i}$.  As will be discussed in detail below, the prefactor $\xi^{\eta}_{ijk}$ only depends on the properties of monolayer graphene, and is of \text{opposite sign} for the opposite valleys. Therefore, the opposite Chern numbers together with opposite strain effects  means that the PET contributed by the two  valleys should be the same, and is exactly quantized in terms of the VCNs of the occupied bands multiplied by a constant:
\begin{equation}
    \gamma_{ijk}=2(-1)^i\frac{e}{2\pi}(\xi^{-}_{ijk}C_{-}+\xi^{+}_{ijk}C_{+}),
\label{eq.gamma_CN}
\end{equation}
where $C_{\pm}$ is the Chern number of the occupied bands from the $K$ and $K'$ valleys, and  $(-1)^i=\pm 1$ for $i=x,y$.
The prefactor $2$ accounts for the spin degeneracy in the moir\'e graphene systems. 
The quantization of the PET can be intuitively interpreted as follows. Let us consider the situation that the strain-induced effective vector potential $\mathbf{A}^{\eta}$ is adiabatically turned on in the system, which imposes opposite effective ``electric fields" $\mathbf{E}^{\eta}=-\dot{\mathbf{A}}^{\eta}$ ($\eta=\pm$) to the two valleys, which in turn induce transverse  adiabatic ``Hall currents" from the two valleys with opposite Chern numbers. The adiabatic ``Hall currents" generated from the two  valleys are the same since both the Chern numbers and the effective electric  fields are opposite for the two valleys. The piezoelectric tensor, e.g., $\gamma_{yxx}$, measures precisely the ``Hall conductivity" of the adiabatic current in response to the opposite effective electric fields applied to the two valleys in the process of adiabatically turning on the strain, and such response is naturally quantized in units of VCNs of the system. 
In a previous work \cite{piezo_TQPT_Natcomm}, it is shown that for topologically non-trivial 2D crystalline materials with TR symmetry, a jump of piezoelectric response is expected near a topological phase transition (TPT).  In the above argument, we show that in moir\'e graphene systems the PET does not only exhibit discontinuity across a TPT, but are always quantized in units of the VCNs of the occupied bands no matter the system is close to a TPT or not. This is a unique property for the topologically nontrivial bands with nonzero VCNs in moir\'e graphene systems by virtue of the valley $U(1)$ symmetry.

\paragraph{Model Hamiltonian \textemdash}
To illustrate the idea of quantized PET, we first study the PET of hBN-aligned TBG based on Bistritzer-MacDonald continuum Hamiltonian \cite{macdonald-pnas11}. The effects of the hBN alignment is modeled by imposing a staggered sublattice potential $\Delta\!=\!15\,$meV to both layers, neglecting the additional moir\'e potential introduced by the lattice match between hBN and graphene \cite{senthil-hbn-tbg-prr19,zaletel-tbg-2019}, which is an order of magnitude weaker than that of TBG \cite{koshino-prb14,jung-prb14}. The aligned hBN substrate breaks $C_{2z}$ symmetry, opening a gap $\sim\,4\,$meV between the conduction and valence bands (see Fig.~\ref{fig1}(a)), which have nonzero VCNs $\pm 1$. The original $D_6$ symmetry of the continuum model is reduced to $D_3$ symmetry, which requires the PET components subject to the following relation:
\begin{align}
    &\gamma_{xxx}=\gamma_{xyy}=\gamma_{yxy}=0 \;\nn
    &\gamma_{xxy}=\gamma_{yxx}=-\gamma_{yyy}
    \label{eq:gamma}
\end{align}
As a result of the $D_3$ symmetry, there is only one independent PET component $\gamma_{yxx}$. Note that Eq.~(\ref{eq:gamma}) holds as long as the system has $C_3$ symmetry, which is present in all TMG systems. Therefore we choose to illustrate our results through the component $\gamma_{yxx}$ for all twisted graphene systems discussed in this work.
The Hamiltonian in the continuum model of TBG in $K_\eta$ ($\eta=\mp$ represents $K$ and $K^\prime$ valleys respectively) is expressed as 
\begin{equation}
    H^\eta(\mathbf{r})=\begin{pmatrix}
        H_1^\eta & U_{\rm M}(\mathbf{r}) \\
        U_{\rm M}^\dagger(\mathbf{r}) & H_2^\eta 
    \end{pmatrix},
\label{eq.HTBG}
\end{equation}
where the diagonal blocks $H_{l}^\eta$ are the monolayer layer graphene (MLG) Dirac cones at $K_\eta$ valley of layer $l$. The off-diagonal block, which represents the interlayer coupling between the twisted bilayers  introduces a moir\'e potential in real space $U_M(\mathbf{r})$, and the detailed expression is given in Supplementary Information \cite{supp_info}. 
As mentioned in the previous section, a homogeneous strain applied to the system is equivalent to a pseudo vector potential. Here we ignore the effects of strain on the interlayer hopping \cite{bi-tbg-strain}, which are considered as higher-order effects, then the strain effects are only manifested in the diagonal blocks, which become
\begin{equation}
    H_l^\eta=\hbar v_F\,({\mathbf{q}}+\eta{\mathbf{A}}_{\textrm{MLG}})\cdot (\eta\sigma_x, \sigma_y)+\Delta\sigma_z,
\label{eq.intralayer}
\end{equation}
where ${\mathbf{q}}={\mathbf{k}}-{\mathbf{K}}_{l,\eta}$. For MLG in its $K$ valley, 
\begin{equation}
    {\mathbf{ A}}_{\textrm{MLG}}=-\frac{\sqrt{3}\beta}{2a}(\mu_{xx}-\mu_{yy}, -2\mu_{xy}),
\end{equation}
where $a=2.46\,\angstrom$ is the lattice constant of MLG and $\beta=3.14$ is the decaying rate of the hopping amplitude with respect to the distance between two carbon atoms in MLG \cite{bi-tbg-strain}. This expression of strain induced vector field together with Eq.~(\ref{eq.gamma_CN}) tells us that in TBG system, the PET component $\gamma_{yxx}$ per spin per VCN is 
\begin{equation}
    \gamma_{yxx}^0=\frac{e}{2\pi}\frac{\sqrt{3}\beta}{2a}=-281.8{\rm pC/m}.
\label{eq.gamma0}
\end{equation}
Note that $\gamma_{yxxx}^{0}$ only depends on two paramters: the graphene lattice constant $a$ and the  hopping decaying rate $\beta$, both of which are ``fundamental constants" in the universe of graphene-based systems. In other words, the quantization of the PET in units of $\gamma_{yxx}^0$  remains robust regardless of the change of external tuning parameters such as external fields and twist angles.

\begin{figure}[!htbp]
\includegraphics[width=3.5in]{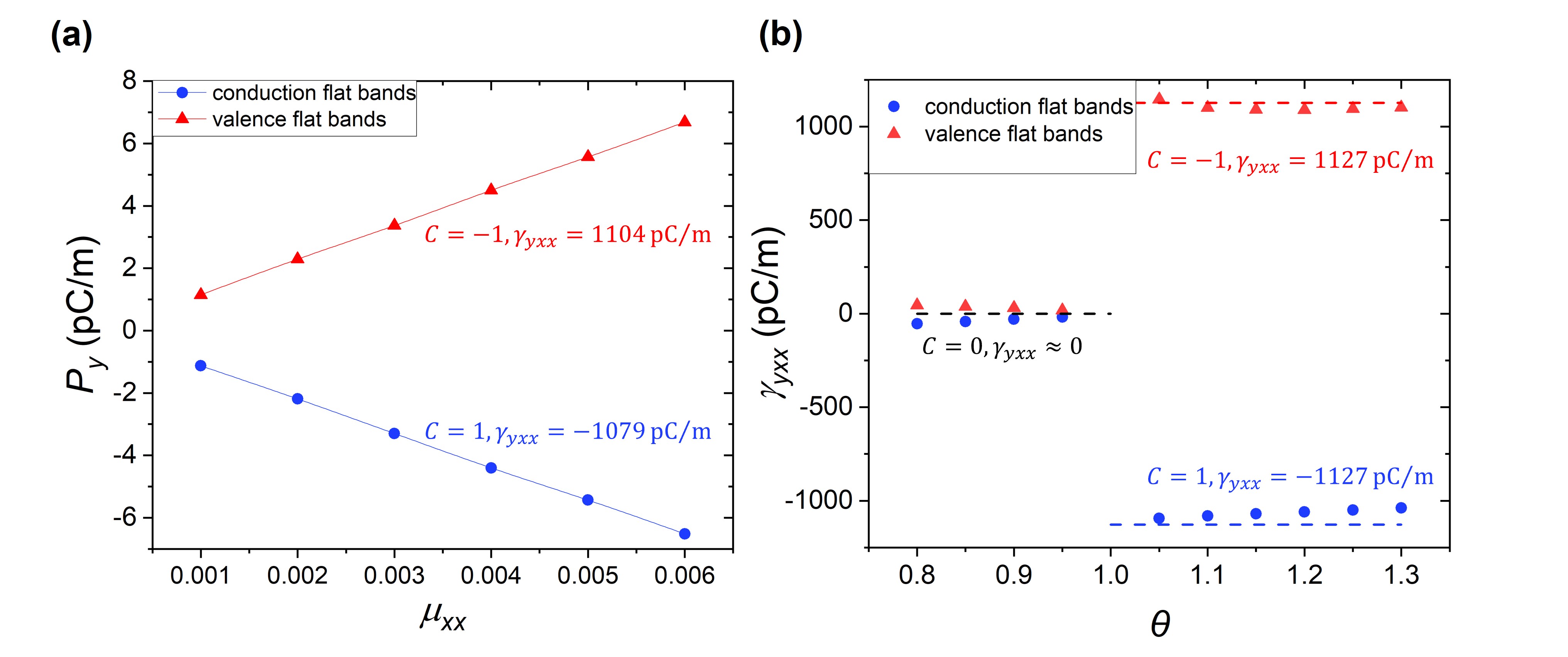}
\caption{~\label{fig2} (a) Plots of polarization along $y$ direction \textit{versus} strain $u_{xx}$ for hBN-aligned TBG at $\theta=1.05\,^{\circ}$, and (b) twist-angle ($\theta$) dependence of $\gamma_{yxx}$ contributed by the valence (red triangles) and conduction (blue dots) flat bands. The horizontal dashed lines mark the ideal quantized values.}
\end{figure}

In Fig.~\ref{fig2}(a) we present the polarization along $y$ direction as a function of strain $u_{xx}$, with twist angle $\theta=1.05\,^{\circ}$ for hBN-aligned TBG, where the blue circles and red triangles represent the contributions from the valence and conduction flat bands with opposite Chern numbers $\pm 1$. Clearly the strain-induced polarizations are opposite from the two flat bands, indicating opposite PETs. In Fig.~\ref{fig2}(b) we show the PET $\gamma_{yxx}$ contributed by the valence (blue circles) and conduction (red triangles) flat bands as a function of twist angle $\theta$. When $\theta\gtrapprox 1^{\circ}$, the VCNs of the two flat bands are $\pm 1$, the calculated $\gamma_{yxx}$ is around the expected quantization value $\pm 4\gamma_{yxx}^{0}$ (marked by the dashed lines), where the prefactor 4 is from the fourfold valley-spin degeneracy. Under the present parameter choice, when $\theta\lessapprox 1^{\circ}$, there is a TPT due to gap closures between the flat bands and the remote bands, such that the VCNs of the flat bands become zeros. As a result, $\gamma_{yxx}$ rapidly drops and becomes negligibly small as shown in Fig.~\ref{fig1}(b). The plateau shape of $\gamma_{yxx}$-$\theta$ relationship supports our argument for the quantized PET. Fig.~\ref{fig2}(b) also suggests a quantized jump of PET by $4\gamma_{yxx}^{0}$  when the carrier density is changed by $\pm 4$ per moir\'e supercell with respect to the charge neutrality point.

\begin{figure}[!htbp]
\includegraphics[width=0.5\textwidth]{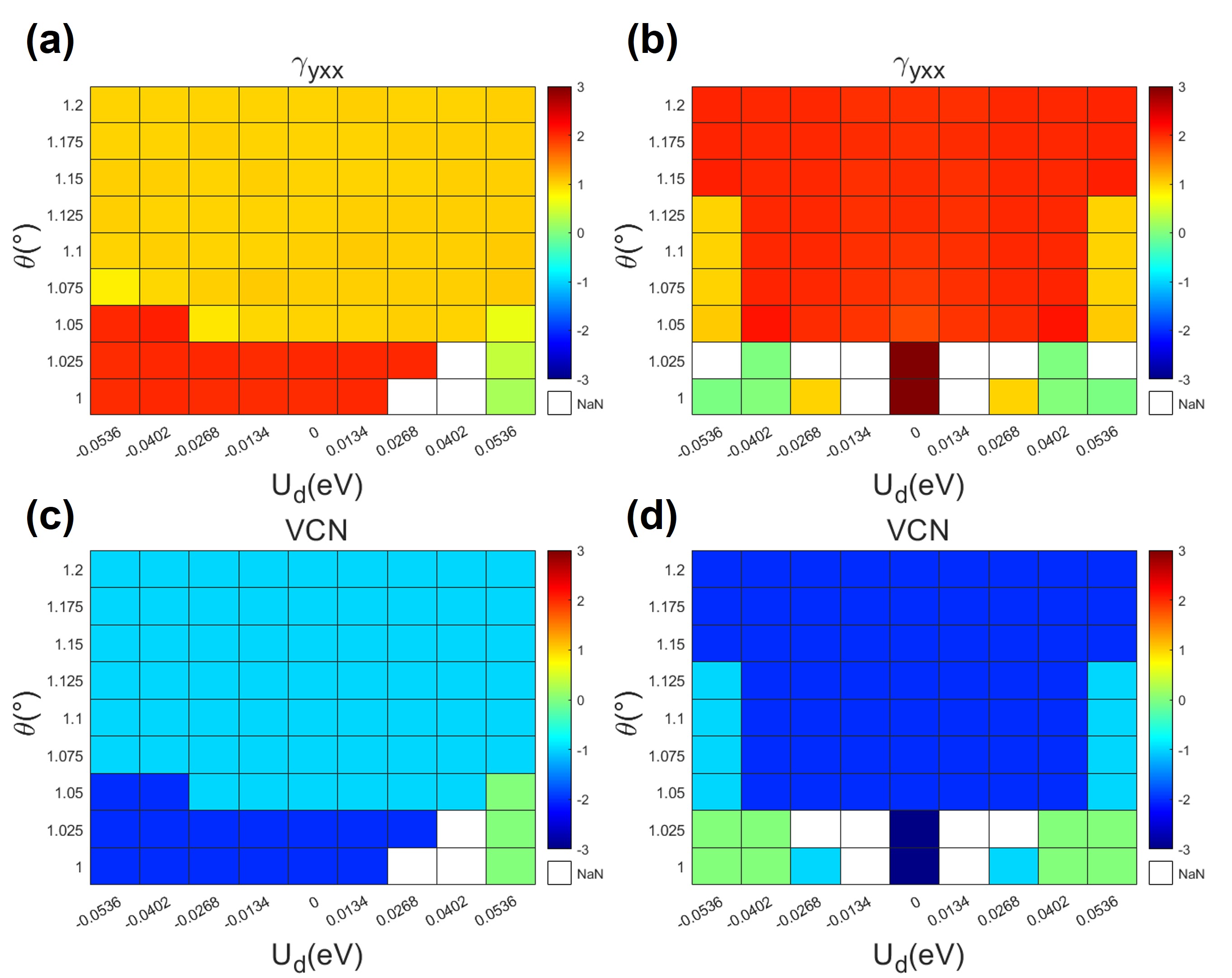}
\caption{~\label{fig3} PET and VCNs of all flat bands in TBMG and TDBG systems: (a) $\gamma_{yxx}$ of TBMG as a function of $\theta$ and $U_d$; (b) $\gamma_{yxx}$ of AB-BA stacked TDBG as a function of $\theta$ and $U_d$; (c) Total Chern numbers of the two flat bands from ${\rm K}$ valley of TBMG as a function of $\theta$ and $U_d$; (d) Total Chern numbers of the two flat bands from ${\rm K}$ valley of AB-BA stacked TDBG as a function of $\theta$ and $U_d$. In Fig.~(a) and (b), the values of $\gamma_{yxx}$ are shown in the units of $|4\gamma_{yxx}^0|=1127{\rm pC/m}$. The blank patches indicate points which are too close to gap closures such that VCNs are ill defined.}
\end{figure}

\paragraph{PET in TMG systems \textemdash}
The continuum model for TBG can be readily extended to TMG systems, for valley $\eta$, the Hamiltonian is expressed as \cite{jpliu-prx19}
\begin{equation}
    H^\eta_{M,\alpha;N,\alpha^\prime}=\begin{pmatrix}
        H^\eta_{M,\alpha} & U \\
        U^\dagger & H^\eta_{N,\alpha^\prime}
    \end{pmatrix}\;,
\label{eq.H_TMG}
\end{equation}
where the diagonal blocks refers to $M(N)$-layer untwisted rhombohedral graphene multilayers with stacking chiralities $\alpha(\alpha^\prime)=\pm$ \cite{supp_info}.  The off-diagonal block of Eq.~(\ref{eq.H_TMG}) is an simple extension of the moir\'e potential in TBG in the sense that we only consider the moir\'e potential  term  between the topmost layer of the $M$ layers and the bottom-most layer of the $N$ layers. We refer readers to Supplementary Information for more details. The band structures for TBMG, AB-BA stacked TDBG, and AB-BA stacked TDBG calculated using the continuum model are presented in Fig.~\ref{fig1}(b)-(d) with Chern numbers of the flat bands from the $K$ valley being marked.
It is worthwhile to note that, with a realistic description of the system including further neighbor interlayer hoppings within each set of the untwisted multilayers \cite{supp_info},  the strain may couple to the interlayer hopping in a similar, but slightly different way from how it couples to the intralayer hopping terms. In particular, the pseudo vector potential associated with further-neighbor interlayer hopping event is expressed as
\begin{equation}
    {\mathbf{A}}_{\textrm{inter}}=-\lambda\frac{\sqrt{3}\beta}{2a}(\mu_{xx}-\mu_{yy}, 2\mu_{xy})=\lambda {\mathbf{A}}_{\textrm{MLG}},
\end{equation}
where $\lambda=a_0/\sqrt{a_0^2+d_{\rm AB}^2}$, with $a_0=a/\sqrt{3}\approx 1.42\,\angstrom$ and $d_{\rm AB}=3.35\,\angstrom$ being the in-plane carbon-carbon $\sigma$-bond length and the interlayer distance between AB-stacked bilayer graphene respectively. As a result of the different strain-momentum coupling terms, the substitution $\partial_{\mu_{jk}}=\xi_{ijk} \partial_{k_i}$ in deriving Eq.~(\ref{eq.gamma_CN}) no longer holds. Luckily, since this quantization-breaking mechanism only occurs in the further-neighbor interlayer hopping terms \cite{supp_info}, for typical TMGs such as TBMG, and TDBG systems, the deviations of PET from the expected quantizated values are very weak (see Fig.~\ref{fig3}). 

In Fig.~\ref{fig3}, we show  $\gamma_{yxx}$ component 
as a function of twist angle $\theta$ and vertical electrostatic potential energy drop across the multilayers, denoted by $U_d$.
Here $U_d$ is related to the displacement field $D$ via: $U_d=-eDd/\varepsilon_{\rm hBN}$, where $d=(M+N-1)\times 3.35\,\angstrom$ is the total thickness of the twist $(M+N)$-layer TMG system, and $\varepsilon_{\rm hBN}\approx 5$ is the dielectric constant of the hBN substrate, \textit{e.g.}, $D=0.1{\rm V/nm}$ corresponds to $U_d=-0.0134\,(M+N-1)\,$eV. 
As suggested by Ref.~\cite{jpliu-prx19}, in the chiral limit, the total VCNs of the two flat bands per spin equals to $\pm 1$ for TBMG and $\pm 2$ for AB-BA stacked TDBG. This conclusion conforms to the calculated values of $\gamma_{yxx}$ and VCNs in the largest plateau of each subplot of Fig.~\ref{fig3}. As the further neighbor interlayer hoppings  and external displacement fields break the chiral limit, the total VCNs of the two flat bands change as a function $U_d$ and $\theta$,  as shown in Fig.~\ref{fig3}(c) and (d) for TBMG and TDBG respectively, hence making the PETs jump to different quantized plateaus as shown in Fig.~\ref{fig3}(a) and (b). The calculated PETs and VCNs agree with each other very well. 

As the flat bands are reported to be isolated from the remote bands for these TMG systems \cite{young-monobi-nature20,Yankowitz-monobi-np2020,Pasupathy-nematic-tdbg-arxiv20,Yankowitz-doublebi-np2020,kim-tdbg-nature20,cao-tdbg-nature20,zhang-tdbg-np20}, we suggest that PET contributed by the flat bands can be extracted in experiments by comparing the PETs with filling factor $+4$ and $-4$, i.e., when the flat bands are fully occupied or empty. By tuning the vertical displacement field, the PET contributed by the flat bands would remain stable and nearly constant, until the field is strong enough to drive a TPT, with a quantized jump of PET. In this way, the VCNs of the flat bands of the TMG systems can be directly detected by measuring the piezoelectric response.

The above argument of quantized PET is based on two approximations: (\i) the $K$ and $K'$ valleys are decoupled, which is an excellent approximation for moir\'e graphene systems and is widely adopted in literatures, and (\ii) the effects of homogeneous strain on the moir\'e potentials have been neglected, which deserves further verification. We have further checked the validity of approximation (\ii) by directly calculating the piezoelectric response of hBN-aligned TBG and TBMG based on a realistic atomistic Slater-Koster tight-binding model adopted from Ref.~\cite{moon-tbg-prb13}, which can include all the strain effects on the electronic structures. 
For hBN-aligned TBG, our results indicate that $\gamma_{yxx}$ \textit{vs.} $\theta$ plot still exhibits a plateau shape with an abrupt jump at $\theta\approx1\,^{\circ}$ due to a TPT (VCNs become zeros for $\theta\lessapprox 1^{\circ}$)  although the quantization is not quite exact due to the strain effects on the moir\'e potential, with $\gamma_{yxx}\approx \pm 1.2\times 4\gamma_{yxx}^{0}$  when the VCNs are $\pm 1$ for $\theta\gtrapprox 1^{\circ}$ \cite{supp_info}. We have also calculated the $U_d$ dependence of $\gamma_{yxx}$ in TBMG contributed by the two flat bands at $\theta=1.25^{\circ}$, and find a plateau of $\gamma_{yxx}\approx 1.15\times 4\gamma_{yxx}^{0}$ \cite{supp_info}, compatible with the total VCN $\pm 1$.

In this work we apply the modern polarization theory for crystals to the case of piezoelectric response in moir\`e graphene systems, and suggest a possible way of measuring the VCNs in these systems based on piezoelectric response. 
We propose that for moir\`e graphene systems with valley charge conservation,  the PET is exactly quantized as integer multiples of the VCNs under certain mild approximations. Although this quantization is not exact in a more realistic situation, it turns out that the quantization condition is only slightly deviated, so that the plateau patterns of the PET are still clearly present for typical twisted graphene systems such as hBN-aligned TBG, TBGM, and TDBG systems. By tuning external displacement fields and twist angle, which can affect the topological properties of the flat bands in the TMG systems, the PET plateaus can be directly measured by experiments, which manifest the VCNs of the flat bands.

\acknowledgements
This work is supported by the National Key R \& D program of China (grant no. 2020YFA0309601), the National Science Foundation of China (grant no. 12174257), and the start-up grant of ShanghaiTech University. We would like to thank Prof. Yue Zhao for valuable discussions. 

\appendix

\section{Continuum model of twisted bilayer graphene systems}

\begin{figure}[!htbp]
\includegraphics[width=2.5in]{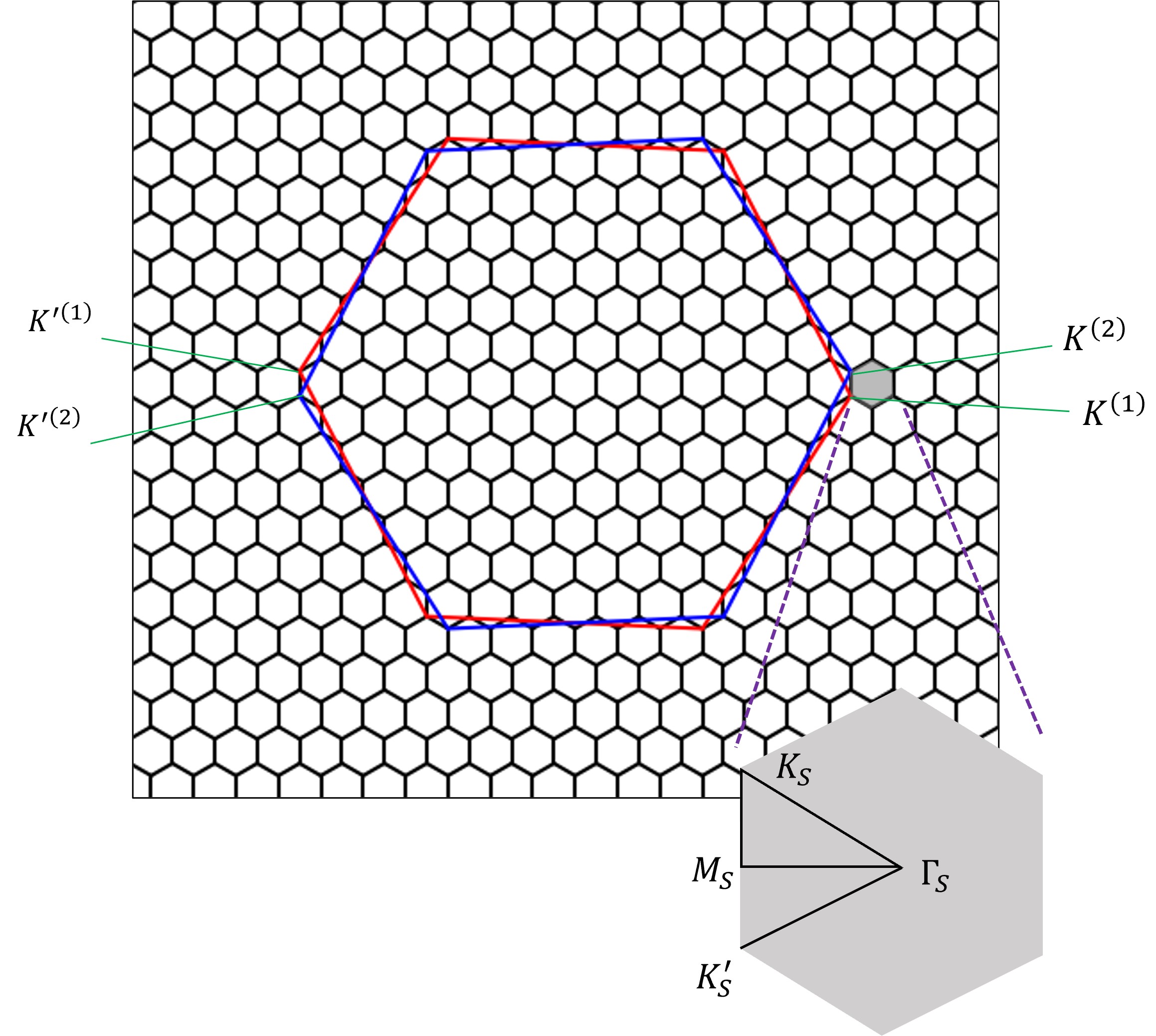}
\caption{~\label{Fig1SM} A plot of the Brillouin zones of the two layers of monolayer graphene (MLG) (where layer 1 is marked with red color and layer 2 is marked with blue color) and the moir\`e Brillouin zone (mBZ) of TBG (each hexagonal patch of the lattice). The inset zooms in a moir\`e Brillouin zone and the high symmetry path used in Fig.~1 of the main text is marked. The twist angle in the plot ($\sim 5^\circ$) is larger than the ones in main text for clarity. }
\end{figure}

To introduce the continuum model of moir\`e graphene systems \cite{macdonald-pnas11, jpliu-prx19}, we first start with twisted bilayer graphene (TBG). The general idea of continuum model is to express the low energy Hamiltonian $H^\eta$ of TBG in the basis of  MLG Bloch states. As shown in Fig.~\ref{Fig1SM}, the moir\'e Brillouin zone (mBZ) of TBG is much smaller than the MLG Brillouin zone for small twist angle $\theta$. As a result, by choosing a mBZ that is near the $K(K^\prime)$ valley of MLG, the intralayer part of TBG Hamiltonian can be well approximated by the Dirac cone form of the MLG Hamiltonian: 
\begin{align}
    &\langle {\mathbf{k}}_l,X_l |H^\eta| {\mathbf{k}}_l,X_l^\prime\rangle=(H^\eta_l)_{X_l X_l^\prime}, \; \nn
    &H^\eta_l=\hbar v_FR(-\theta_l)\,(\mathbf{q}_l+\eta \mathbf{A}_{\rm MLG})\cdot (\eta\sigma_x, \sigma_y), 
\label{eq.Dirac_cone}
\end{align}
where $\eta=\pm$ refers to $K(K^\prime)$ valley, ${\mathbf{q}}_l={\mathbf{k}}_l-{\mathbf{K}}_{l,\eta}$ is the momentum relative to the $K_\eta$ valley of layer $l$, $X_l,X_l^\prime=A_l,B_l$ refers to the $A$, $B$ sublattices of each layer, and $\mathbf{A}_{\rm MLG}$ is the strain induced vector potential (see main text). If strain and staggered sublattice potential are included, as in main text, we should add an extra mass term $\Delta \sigma_z$ to the second equation of Eq.~(\ref{eq.Dirac_cone}) (see main text). Since the twist angles are generally very small in such systems, we can take the rotational operation $R(\theta_l)$ to be identity. We choose $\hbar v_F/a=2.1354\,{\rm eV}$ as in Ref.~(\onlinecite{MLWF-1}), where $a=2.46\,\angstrom{}$ is the lattice constant of MLG. 

To consider the interlayer part of $H^\eta$, we start with an untwisted AA stacked bilayer graphene. The two layers are then twisted by $\theta_l=(-1)^l\theta/2$ around a common $A$ sublattice site. The general form of an interlayer term for hopping from layer 2 to layer 1 is 
\begin{align}
    \langle {\mathbf{k}}_1,X_1 |H^\eta| {\mathbf{k}}_2,X_2\rangle=\frac{1}{\sqrt{N_1 N_2}}\sum_{{\mathbf{R}}_1,{\mathbf{R}}_2} e^{-i{\mathbf{k}}_1\cdot({\mathbf{R}}_1+{\boldsymbol{\tau}}_{1,X_1})} \; \nn
    \times t_{12}^{X_1 X_2}({\mathbf{R}}_1,{\mathbf{R}}_2)e^{i{\mathbf{k}}_2\cdot({\mathbf{R}}_2+{\boldsymbol{\tau}}_{2,X_2})},
\label{eq.interlayer1}
\end{align}
where ${\mathbf{R}}_l$ are the lattice points of the two layers, and $t_{12}^{X_1 X_2}=t({\mathbf{R}}_1+{\boldsymbol{\tau}}_{1,X_1}-{\mathbf{R}}_2-{\boldsymbol{\tau}}_{2,X_2})$ is the hopping amplitude between two atomic orbitals, which has the Slater-Koster form \cite{moon-tbg-prb13}:
\begin{align}
    t({\mathbf{R}})&=V_{pp\pi}[1-(\frac{{\mathbf{R}}\cdot {\mathbf{e}}_z}{R})^2] + V_{pp\sigma}(\frac{{\mathbf{R}}\cdot {\mathbf{e}}_z}{R})^2, \; \nn
    V_{pp\pi}&=V_{pp\pi}^0 e^{-(R-a_0)/r_0}, \qquad V_{pp\sigma}=V_{pp\sigma}^0 e^{-(R-d_0)/r_0},
\label{eq.Slater-Koster}
\end{align}
where $a_0=a/\sqrt{3}$ is the in-plane carbon-carbon sigma bond length, and $d_0=d_{\rm AB}=3.35\,\angstrom{}$ is the interlayer distance between AB-stacked bilayer graphene. The parameters in Eq.~(\ref{eq.Slater-Koster}) are chosen to be: $V_{pp\pi}^0=-2.7\,{\rm eV}$, $V_{pp\sigma}^0=0.48\,{\rm eV}$, and $r_0=0.184a$ so as to fit the hopping within MLG and between AB-stacked bilayer graphene. Note that for in-plane hopping, Eq.~(\ref{eq.Slater-Koster}) is reduced to 
\begin{equation}
    t(R)=V_{pp\pi}^0 e^{-\beta(R/a_0-1)},
\label{eq.in-plane_SK}
\end{equation}
where $\beta=a_0/r_0=3.14$ is the decaying rate that appears in the strain induced vector field in graphene (see main text). By Fourier transforming the hopping amplitude in Eq.~(\ref{eq.interlayer1}), the interlayer hopping term finally becomes (see more details in Ref.~\onlinecite{moon-tbg-prb13,koshino-prx18})
\begin{equation}
    \langle {\mathbf{k}}_1,X_1 |H^\eta| {\mathbf{k}}_2,X_2\rangle=\frac{1}{S_{\rm M}}\int_{S_{\rm M}} {\rm d}^2{\mathbf{r}}\,(U_{\rm M})_{X_1 X_2}\,e^{i({\mathbf{k}}_1-{\mathbf{k}}_2)\cdot {\mathbf{r}}}.
\end{equation}
The Hamiltonian $H^\eta$ now can be written under the basis ${|A_1\rangle,|B_1\rangle,|A_2\rangle,|B_2\rangle}$ as 
\begin{equation}
    H^\eta(\mathbf{r})=\begin{pmatrix}
        H_1^\eta & U_{\rm M}(\mathbf{r}) \\
        U_{\rm M}^\dagger(\mathbf{r}) & H_2^\eta
    \end{pmatrix},
\end{equation}
where 
\begin{equation}
\begin{split}
    U_{\rm M}(\mathbf{r})=\begin{pmatrix}
        u & u^\prime \\
        u^\prime& u
    \end{pmatrix}
    +\begin{pmatrix}
        u & u^\prime \omega^{-\eta} \\
        u^\prime \omega^{\eta} & u
    \end{pmatrix} e^{i\eta{\mathbf{G^M_2}}\cdot{\mathbf{r}}} \\
    +\begin{pmatrix}
        u & u^\prime \omega^{\eta} \\
        u^\prime \omega^{-\eta} & u
    \end{pmatrix} e^{-i\eta{\mathbf{G^M_1}}\cdot{\mathbf{r}}},
\end{split}
\label{eq.U_twist}
\end{equation}
with $\omega=e^{i2\pi/3}$, $\mathbf{G^M_1}=4\pi/(\sqrt{3}L_s)(1/2,-\sqrt{3}/2)$, $\mathbf{G^M_2}=4\pi/(\sqrt{3}L_s)(1/2,\sqrt{3}/2)$ being the Moir\`e reciprocal basis vectors, and $L_s$ being the moir\`e supercell lattice constant. The parameters are chosen to be $u=0.0797\,{\rm eV}$, $u^\prime=0.0975\,{\rm eV}$, and the difference between $u$ and $u^\prime$ characterizes the corrugation of TBG \cite{koshino-prx18}, whose main effect is to separate the flat bands from the others. The derivation of Eq.~(\ref{eq.U_twist}) is the same as in Ref.~\onlinecite{moon-tbg-prb13}, except for a different choice of basis vectors and sublattices.

\section{Continuum model of twisted multilayer graphene systems}
The continuum model of TBG constructed above can be easily extended to twisted multilayer graphene (TMG) systems. The Hamiltonian of an $(M+N)$-layer TMG system with stacking chiralities $\alpha,\alpha^\prime=\pm$ is expressed as
\begin{equation}
    H^\eta_{M,\alpha;N,\alpha^\prime}=\begin{pmatrix}
        H^\eta_{M,\alpha} & U \\
        U^\dagger & H^\eta_{N,\alpha^\prime}
    \end{pmatrix}\;,
\label{eq.H_TMG}
\end{equation}
where a diagonal block, \textit{e.g.}, for the bottom $M$ layers and positive stacking chirality, is
\renewcommand*{\arraystretch}{1.5}
\begin{equation}
    H^\eta_{M,+}=\begin{pmatrix}
        H_1^\eta & U_{\rm AB} & 0 & 0 & \cdots \\
        U_{\rm AB}^\dagger & H_1^\eta & U_{\rm AB} & 0 & \cdots \\
        0 & U_{\rm AB}^\dagger & H_1^\eta & U_{\rm AB} & \cdots \\
        \vdots & \vdots & \vdots & \vdots & \ddots 
    \end{pmatrix}.
\label{eq.H_M}
\end{equation}
In Eq.~(\ref{eq.H_M}), the diagonal blocks, \textit{i.e.}, the intralayer Hamiltonian, are the same as the one defined in Eq.~(\ref{eq.Dirac_cone}), except that $\Delta=0$ here; and matrix $U_{\rm AB}$ refers to the interlayer hopping between AB stacking bilayer graphene, with 
\renewcommand*{\arraystretch}{1}
\begin{equation}
    U_{\rm AB}=\begin{pmatrix}
        0 & 0 \\
        t_1 & 0
    \end{pmatrix},
\label{eq.UAB_simple}
\end{equation}
where $t_1=0.48{\rm eV}$ is the hopping amplitude between two atoms in different layers with the same horizontal position. If the stacking chirality is reversed in Eq.~(\ref{eq.H_M}), we may simply replace $U_{\rm AB}$ with $U_{\rm BA}=U_{\rm AB}^\dagger$. The off-diagonal block of Eq.~(\ref{eq.H_TMG}) is 
\renewcommand*{\arraystretch}{1.5}
\begin{equation}
    U=\begin{pmatrix}
        0 & 0 & \cdots & 0 \\
        0 & 0 & \cdots & 0 \\
        \vdots & \vdots & \ddots & \vdots \\
        U_{\rm M} & 0 & \cdots & 0
    \end{pmatrix},
\end{equation}
where the only non-zero block refers to the interlayer coupling between the bottom layer of the upper $N$ layers and the top layer of the bottom $M$ layers, which has the same definition as Eq.~(\ref{eq.U_twist}). As discussed in main text, the moir\`e graphene systems should show quantized piezoelectric response as long as the strain linearly couples to the momentum, which is the case of Eq.~(\ref{eq.Dirac_cone}). Under the approximation where the effects by strain to interlayer hopping are ignored, piezoelectricity contributed by low energy bands of TBG is quantized. We can see that under the current model Hamiltonian for TMG, strain still enters the Hamiltonian by linearly coupling to the momentum through vector potential ${\mathbf{A}}_{\rm MLG}$ because Eq.~(\ref{eq.UAB_simple}) is independent of strain. Therefore, our argument for quantized piezeoelectric response in TBG is also valid for the TMG systems. 

\begin{figure}[!htbp]
\includegraphics[width=3.5in]{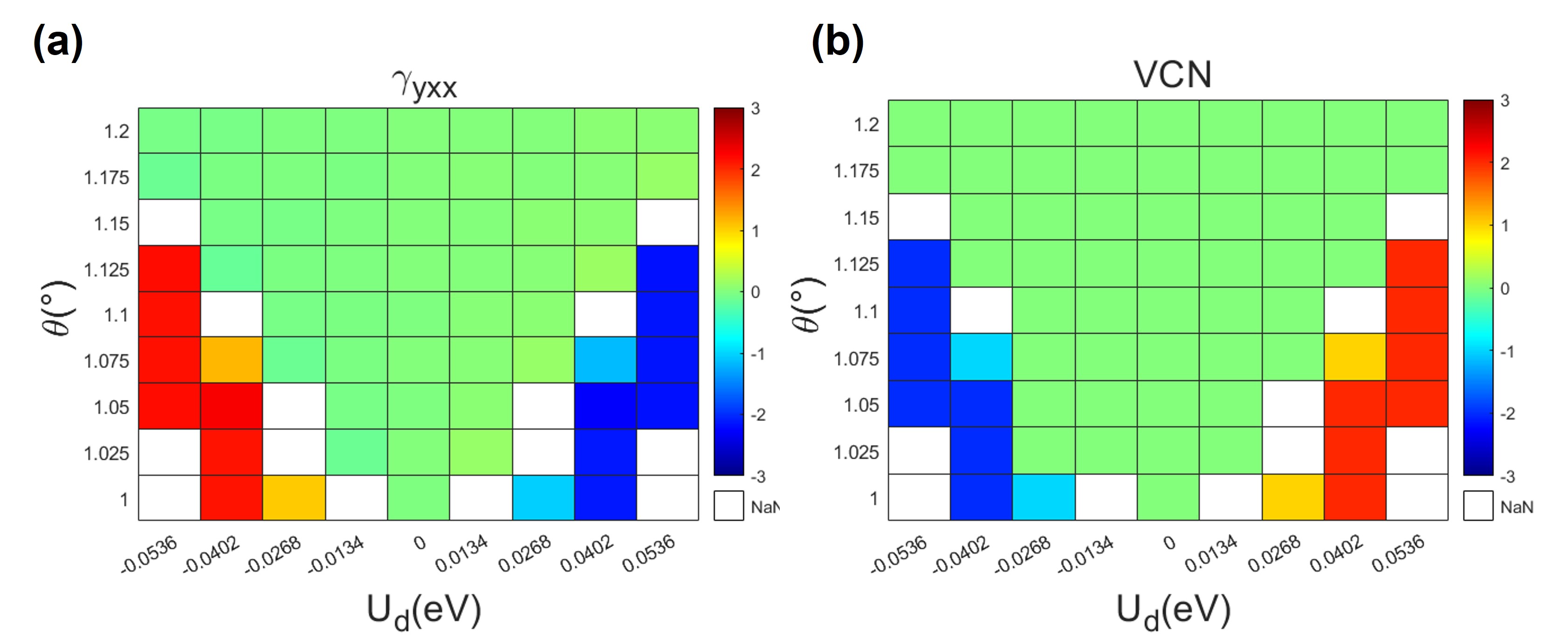}
\caption{~\label{Fig2SM} A plot of (a) the $\gamma_{yxx}$ component of PET and (b) the VCN from $K$ valley of AB-AB stacked TDBG as a function of twist angle $\theta$ and vertical electrostatic potential energy drop $U_d$ across the multilayers.}
\end{figure}

However, In our numerical calculation, we adopt a more realistic version of the interlayer hopping between untwisted layers:
\renewcommand*{\arraystretch}{1}
\begin{equation}
    U_{\rm AB}=\begin{pmatrix}
        t_2f({\mathbf{k}}) & t_2f^*({\mathbf{k}}) \\
        t_1-t_3 & t_2f({\mathbf{k}})
    \end{pmatrix},
\label{eq.UAB_real}
\end{equation}
where $t_2=0.21{\rm eV}$ and $t_3=0.05{\rm eV}$ are the second and third nearest interlayer hopping amplitude, and $f({\mathbf{k}})=-(\sqrt{3}a/2)(\eta q_x-iq_y)$ for momenta near the $K_\eta$ valley. Here 
\begin{equation}
    (q_x,q_y)={\mathbf{k-K}}_{l,\eta}+\eta \mathbf{A}_{\rm inter},
\end{equation}
with $\mathbf{A}_{\rm inter}=\lambda {\mathbf{A}}_{\textrm{MLG}}$, $\lambda=a_0/\sqrt{a_0^2+d_{\rm AB}^2}$, as shown in main text. Now the untwisted interlayer hopping and intralayer hopping couple to the strain in different ways with further-neighbor interlayer hopping, so the piezoelectric response quantization is no longer conserved. Luckily, this quantization breaking mechanism is weak, as is shown by the results in main text. In this part, we also show piezoelectric tensor (PET) of AB-AB stacked twisted double bilayer graphene (TDBG), which is not shown in the main text because in the major area of the parameter space, the valley Chern number (VCN) of all flat bands equal to zero \cite{jpliu-prx19}, and thus the PET by the two flat  bands (per spin per valley) are hard to be measured.

\section{Tight-binding model}
\begin{figure}[!htbp]
\includegraphics[width=2.0in]{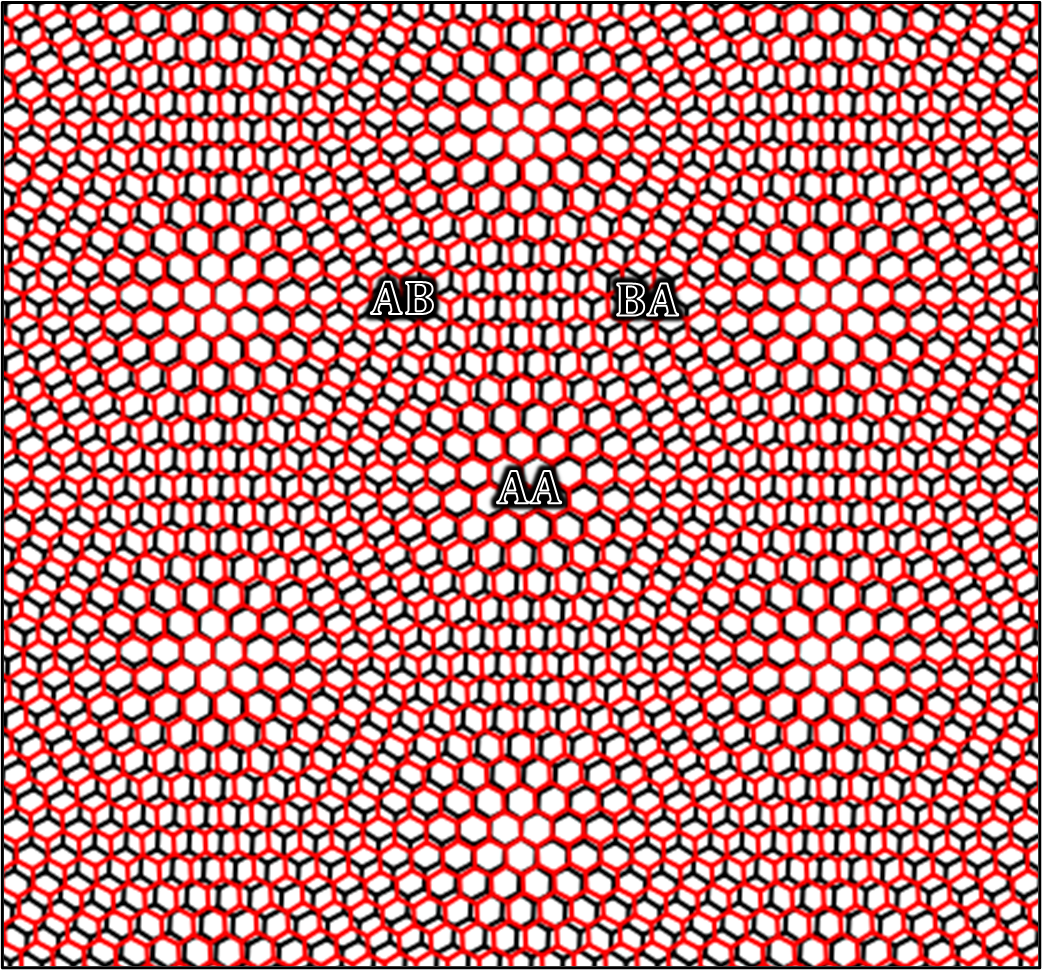}
\caption{~\label{Fig_real_space_moire} A plot of the real space moir\`e pattern in TBG, where the AB, BA, and AA zones are marked.}
\end{figure}
To check the validity of the results based on continuum model, we also implemented atomistic tight-binding model based calculations, which are computationally more demanding, but provide more reliable results. To construct the tight-binding model Hamiltonian, we need to describe the positions of the carbon atoms in a moir\`e supercell. We start with AA stacked bilayer graphene, where the positions of the atoms in layer $l$ $(l=1,2)$ at sublattice $\alpha$ $(\alpha=A,N)$ are given by
\begin{equation}
\begin{split}
    &\mathbf{R}_{l,\alpha}^{mn}=(\mathbf{R}_{l,\alpha}^{mn})_{/\kern -0.6em /}+\mathbf{d}_{l,\alpha}^{mn}, \\
    &(\mathbf{R}_{l,\alpha}^{mn})_{/\kern -0.6em /}=m\mathbf{a}_1+n\mathbf{a}_2+\boldsymbol{\tau}_{\alpha},
\end{split}
\end{equation}
with $\mathbf{a}_1,\mathbf{a}_2$ being the real space basis vectors of graphene, $\boldsymbol{\tau}_{\alpha}$ being the position of a sublattice relative to the graphene unit cell, and $\mathbf{d}_{l,\alpha}^{mn}=(-1)^ld[(\mathbf{R}_{l,\alpha}^{mn})_{/\kern -0.6em /}]/2\,\mathbf{e}_z$ being the vertical displacement of two aligned atoms. After the two layers are twisted by $\theta_l=(-1)^l\theta/2$, the positions of the atoms become $\mathbf{R}_{l,\alpha}^{mn}(\theta_l)=R(\theta_l)\mathbf{R}_{l,\alpha}^{mn}$. As shown in Fig.~\ref{Fig_real_space_moire}, the TBG is divided into AB and AA stacked zones, which share the same period with the moir\`e supercell. As a result, the interlayer distances are different in these zones, \textit{i.e.}, TBG is corrugated. The corrugation can be interpolated by the following function \cite{koshino-prx18}:
\begin{equation}
    d_{l,\alpha}^{mn}=d_0+2d_1 \sum_{i=1}^3 {\rm cos}\mathbf{b}_i\boldsymbol{\delta}[(\mathbf{R}_{l,\alpha}^{mn})_{/\kern -0.6em /}],
\label{eq.interpolated_d}
\end{equation}
where $\mathbf{b}_1=(2\pi/(\sqrt{3}a), 2\pi/a), \mathbf{b}_1=(2\pi/(\sqrt{3}a), -2\pi/a)$ are the reciprocal basis vectors of MLG, and $\mathbf{b}_3=-\mathbf{b}_1-\mathbf{b}_2$. The function $\boldsymbol{\delta}$ is defined as $\boldsymbol{\delta}(\mathbf{R})=[R(\theta_2)-R(\theta_1)]\mathbf{R}$, which refers to the displacement of two aligned atoms in AA stacked bilayer graphene after twisted. In a commensurate structure of TBG, once $\mathbf{R}$ is changed by a moir\`e lattice vector, $\boldsymbol{\delta}(\mathbf{R})$ is changed by an MLG lattice vector, so Eq.~(\ref{eq.interpolated_d}) shares the same period as the moir\`e superlattice. To fit the interlayer distances of $d_{\rm AB}=3.35\angstrom{}$ and $d_{\rm AA}=3.60\angstrom{}$ \cite{graphite-AA,uchida-tbg-prb14}, the parameters are chosen to be $d_0=(d_{\rm AA}+2d_{\rm AB})/3$ and $d_1=(d_{\rm AA}-d_{\rm AB})/9$. 

Once the positions of the atoms in the supercell are determined, it is straight-forward to derive the tight-binding model Hamiltonian with the hopping amplitudes between each pair of atomic orbitals given by Eq.~(\ref{eq.Slater-Koster}). The Hamiltonian under the Bloch basis of TBG is 
\begin{equation}
    H=\sum_{\mathbf{k,R}}\sum_{i,j} t(\mathbf{R}+\boldsymbol{\tau}_i-\boldsymbol{\tau}_j)e^{i\mathbf{k}(\mathbf{R}+\boldsymbol{\tau}_i-\boldsymbol{\tau}_j)}|\mathbf{k},i\rangle \langle \mathbf{k},j|,
\end{equation}
where $\mathbf{R}$ refers to the moir\`e lattice vectors, and $i,j$ represent the sublattices in moir\`e supercell. 

\begin{figure}[!htbp]
\includegraphics[width=3.5in]{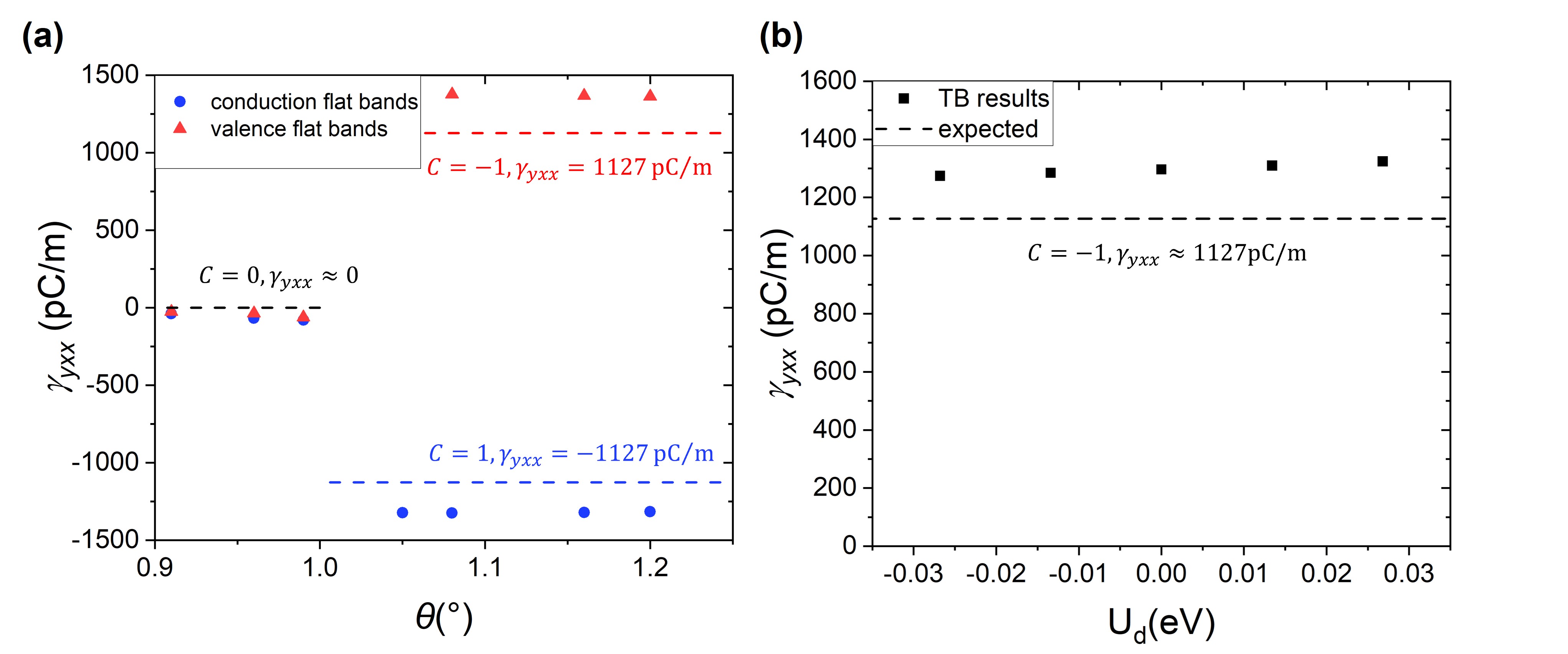}
\caption{~\label{Fig.TB_results} Tight-binding model based results for PET component $\gamma_{yxx}$ of (a) hBN-aligned TBG, contributed by the conduction (blue dots) and valence (red triangles) flat bands, as functions of twist angle $\theta$; and (b) AB-A stacked TMG, contributed by all the flat bands, as a function of vertical electrostatic potential energy drop $U_d$. In both (a) and (b), the predicted values of $\gamma_{yxx}$ for the corresponding bands given by $4\gamma_{yxx}^0=-1127{\rm pC/m}$ per VCN are marked by the dashed lines.}
\end{figure}

In Fig.~(\ref{Fig.TB_results}), we show the calculated $\gamma_{yxx}$ component of PET based on the above atomistic tight-binding model. In particular, in Fig.~(\ref{Fig.TB_results})(a) we show the twist angle dependence of $\gamma_{yxx}$ for hBN-aligned TBG, where the blue circles and red triangles represent $\gamma_{yxx}$ contributed by the valence flat band and conduction flat band with opposite valley Chern numbers $\pm 1$. Clearly we see a plateau shape of $\gamma_{yxx}$ as a function of $\theta$, with an abrupt drop when $\theta\lessapprox 1^{\circ}$. This is because with our choice of parameters, there is an topological phase transition with gap closures between the flat bands and the remote bands, such that the valley Chern numbers of the flat bands become zeros when $\theta\lessapprox 1^{\circ}$. When the valence chern numbers are $\pm 1$, we see that $\gamma_{yxx}\approx\pm 1.2\times(4\gamma_{yxx}^{0})$, which are mildly deviated from the expected quantized value $\pm 4\gamma_{yxx}^{0}$. However, the plateau shape is well preserved, indicating the topological nature of the piezoelectric response. In  Fig.~\ref{Fig.TB_results}(b) we show the calculated $\gamma_{yxx}$ as a function of vertical electrostatic potential drop $U_d$, contributed by the 8 flat bands (including valley and spin degrees of freedom)  in AB-A stacked twisted bilayer-monolayer graphene system with $\theta=1.25^{\circ}$. The total valley Chern number of the two flat bands per spin per valley remains as $\pm 1$ for $-0.03\,\textrm{eV}\leq U_d\leq0.03\,$eV , thus we expect to see  a plateau of $\gamma_{yxx}$ with quantized value of $\pm 4\gamma_{yxx}^{0}$. The calculated $\gamma_{yxx}\approx \pm 1.15\times(4\gamma_{yxx}^{0})$, which still exhibits a plateau shape. The deviations to the expected quantized value are attributed to the strain effects on the moir\'e potentials, which are neglected in continnu model, but captured in the atomistic tight-binding model.    

\section{Derivation of the vector field induced by strain}
We start with the MLG Hamiltonian without approximations:
\begin{equation}
    H_{\rm MLG}=\begin{pmatrix}
        0 & g(\mathbf{k},\hat{\mathbf{\mu}}) \\
        g^*(\mathbf{k},\hat{\mathbf{\mu}}) & 0
    \end{pmatrix},
\end{equation}
where $g(\mathbf{k},\hat{\mathbf{\mu}})=\sum_{i=1}^3 t_0^{(i)}e^{i\mathbf{k}\cdot\mathbf{r}_i}$, with $t_0^{(i)}=t(|\mathbf{r}_i|)$ being the nearest-neighbor hopping amplitude of graphene, $\mathbf{r}_1$, $\mathbf{r}_2$, $\mathbf{r}_3$ being the nearest-neighbor hopping vectors of graphene, and $t(\mathbf{R})$ being the Slater-Koster form defined in Eq.~(\ref{eq.in-plane_SK}). When the system is free of strain, $\mathbf{r}_1^0=(0,a_0)$, $\mathbf{r}_2^0=a_0(\sqrt{3}/2,-1/2)$, $\mathbf{r}_3^0=a_0(-\sqrt{3}/2,-1/2)$. Under presence of strain $\hat{\mathbf{\mu}}$, which has the form of 
\begin{equation}
    \hat{\mathbf{\mu}}=\begin{pmatrix}
        \mu_{xx} & \mu_{xy} \\
        \mu_{xy} & \mu_{yy}
    \end{pmatrix},
\end{equation}
each vector $\mathbf{r}_i$ undergoes small shifts from $\mathbf{r}_i^0$ to $(1+\hat{\mathbf{\mu}})\mathbf{r}_i^0$, which leads to a change $\delta t^{(i)}=t(\mathbf{r}_i)-t(\mathbf{r}_i^0)$ in the hopping amplitudes, \textit{e.g.}, 
\begin{equation}
\begin{split}
    &\delta t(\mathbf{r}_1) \approx -t_0\beta \mu_{yy}, \\
    &\delta t(\mathbf{r}_2) \approx -t_0\beta(\frac{3}{4}\mu_{xx}-\frac{\sqrt{3}}{2}\mu_{xy}+\frac{1}{4}\mu_{yy}), \\
    &\delta t(\mathbf{r}_2) \approx -t_0\beta(\frac{3}{4}\mu_{xx}+\frac{\sqrt{3}}{2}\mu_{xy}+\frac{1}{4}\mu_{yy}),
\end{split}
\end{equation}
where $t_0=t(\mathbf{r}_i^0)$ is the unstrained nearest-neighbor hopping amplitude. By expanding $g(\mathbf{k})$ near the $K_\eta$ valleys, with $\mathbf{K}_\eta=\eta(4\pi/(3a),0)$, we have 
\begin{equation}
    g(\mathbf{q},\hat{\mathbf{\mu}})=g(\mathbf{q},0)+\delta g(\mathbf{q},\hat{\mathbf{\mu}}),
\end{equation}
where $\mathbf{q}=\mathbf{k}-\mathbf{K}_\eta$, and $g(\mathbf{q},0)$ is the off-diagonal component of unstrained Dirac cone:
\begin{equation}
    g(\mathbf{q},0)=\hbar v_F (\eta q_x-iq_y),
\end{equation}
and 
\begin{equation}
    \delta g(\mathbf{q},\hat{\mathbf{\mu}})=-\beta t_0 [(\frac{3}{4}\mu_{xx}-\frac{3}{4}\mu_{yy})+i\eta \frac{3}{2}\mu_{xy}].
\end{equation}
So the expression of $g(\mathbf{q},\hat{\mathbf{\mu}})$ can finally be written as 
\begin{equation}
    g(\mathbf{q},\hat{\mathbf{\mu}})=\hbar v_F [\eta(q_x+\eta A_{\rm MLG}^x)-i(q_y+\eta A_{\rm MLG}^y)],
\end{equation}
or, equivalently,
\begin{equation}
    g(\mathbf{q},\hat{\mathbf{\mu}})=g(\mathbf{q+\eta A_{\rm MLG}},0),
\label{eq.M_to_A}
\end{equation}
with 
\begin{equation}
    A_{\rm MLG}^x=-\frac{\sqrt{3}\beta}{2a}(\mu_{xx}-\mu_{yy}), \qquad
    A_{\rm MLG}^y=-\frac{\sqrt{3}\beta}{a}\mu_{xy}.
\end{equation}
From Eq.~(\ref{eq.M_to_A}) we can see that the effects by strain near the $K_\eta$ valleys of graphene is equivalent to a vector potential. 

The strain induced vector field $\mathbf{A}_{\rm inter}$ (see text) for the further-neighbor hopping between AB stacked bilayers can be derived in similar way, except that the strain only changes the horizontal components of the interlayer carbon-carbon displacement vector. As a result, their distance is changed in a lower rate, and so is the hopping amplitude, which leads to a constant factor $\lambda=a_0/\sqrt{a_0^2+d_{\rm AB}^2}<1$ to the corresponding vector potential.

\bibliography{tmg}

\begin{thebibliography}{52}
\expandafter\ifx\csname natexlab\endcsname\relax\def\natexlab#1{#1}\fi
\expandafter\ifx\csname bibnamefont\endcsname\relax
  \def\bibnamefont#1{#1}\fi
\expandafter\ifx\csname bibfnamefont\endcsname\relax
  \def\bibfnamefont#1{#1}\fi
\expandafter\ifx\csname citenamefont\endcsname\relax
  \def\citenamefont#1{#1}\fi
\expandafter\ifx\csname url\endcsname\relax
  \def\url#1{\texttt{#1}}\fi
\expandafter\ifx\csname urlprefix\endcsname\relax\def\urlprefix{URL }\fi
\providecommand{\bibinfo}[2]{#2}
\providecommand{\eprint}[2][]{\url{#2}}

\bibitem[{\citenamefont{Cao et~al.}(2018{\natexlab{a}})\citenamefont{Cao,
  Fatemi, Fang, Watanabe, Taniguchi, Kaxiras, and
  Jarillo-Herrero}}]{cao-nature18-supercond}
\bibinfo{author}{\bibfnamefont{Y.}~\bibnamefont{Cao}},
  \bibinfo{author}{\bibfnamefont{V.}~\bibnamefont{Fatemi}},
  \bibinfo{author}{\bibfnamefont{S.}~\bibnamefont{Fang}},
  \bibinfo{author}{\bibfnamefont{K.}~\bibnamefont{Watanabe}},
  \bibinfo{author}{\bibfnamefont{T.}~\bibnamefont{Taniguchi}},
  \bibinfo{author}{\bibfnamefont{E.}~\bibnamefont{Kaxiras}}, \bibnamefont{and}
  \bibinfo{author}{\bibfnamefont{P.}~\bibnamefont{Jarillo-Herrero}},
  \bibinfo{journal}{Nature} \textbf{\bibinfo{volume}{556}}, \bibinfo{pages}{43}
  (\bibinfo{year}{2018}{\natexlab{a}}).

\bibitem[{\citenamefont{Yankowitz et~al.}(2019)\citenamefont{Yankowitz, Chen,
  Polshyn, Zhang, Watanabe, Taniguchi, Graf, Young, and
  Dean}}]{dean-tbg-science19}
\bibinfo{author}{\bibfnamefont{M.}~\bibnamefont{Yankowitz}},
  \bibinfo{author}{\bibfnamefont{S.}~\bibnamefont{Chen}},
  \bibinfo{author}{\bibfnamefont{H.}~\bibnamefont{Polshyn}},
  \bibinfo{author}{\bibfnamefont{Y.}~\bibnamefont{Zhang}},
  \bibinfo{author}{\bibfnamefont{K.}~\bibnamefont{Watanabe}},
  \bibinfo{author}{\bibfnamefont{T.}~\bibnamefont{Taniguchi}},
  \bibinfo{author}{\bibfnamefont{D.}~\bibnamefont{Graf}},
  \bibinfo{author}{\bibfnamefont{A.~F.} \bibnamefont{Young}}, \bibnamefont{and}
  \bibinfo{author}{\bibfnamefont{C.~R.} \bibnamefont{Dean}},
  \bibinfo{journal}{Science} \textbf{\bibinfo{volume}{363}},
  \bibinfo{pages}{1059} (\bibinfo{year}{2019}).

\bibitem[{\citenamefont{Codecido et~al.}(2019)\citenamefont{Codecido, Wang,
  Koester, Che, Tian, Lv, Tran, Watanabe, Taniguchi, Zhang
  et~al.}}]{marc-tbg-19}
\bibinfo{author}{\bibfnamefont{E.}~\bibnamefont{Codecido}},
  \bibinfo{author}{\bibfnamefont{Q.}~\bibnamefont{Wang}},
  \bibinfo{author}{\bibfnamefont{R.}~\bibnamefont{Koester}},
  \bibinfo{author}{\bibfnamefont{S.}~\bibnamefont{Che}},
  \bibinfo{author}{\bibfnamefont{H.}~\bibnamefont{Tian}},
  \bibinfo{author}{\bibfnamefont{R.}~\bibnamefont{Lv}},
  \bibinfo{author}{\bibfnamefont{S.}~\bibnamefont{Tran}},
  \bibinfo{author}{\bibfnamefont{K.}~\bibnamefont{Watanabe}},
  \bibinfo{author}{\bibfnamefont{T.}~\bibnamefont{Taniguchi}},
  \bibinfo{author}{\bibfnamefont{F.}~\bibnamefont{Zhang}},
  \bibnamefont{et~al.}, \bibinfo{journal}{Science Advances}
  \textbf{\bibinfo{volume}{5}} (\bibinfo{year}{2019}).

\bibitem[{\citenamefont{Lu et~al.}(2019)\citenamefont{Lu, Stepanov, Yang, Xie,
  Aamir, Das, Urgell, Watanabe, Taniguchi, Zhang et~al.}}]{efetov-nature19}
\bibinfo{author}{\bibfnamefont{X.}~\bibnamefont{Lu}},
  \bibinfo{author}{\bibfnamefont{P.}~\bibnamefont{Stepanov}},
  \bibinfo{author}{\bibfnamefont{W.}~\bibnamefont{Yang}},
  \bibinfo{author}{\bibfnamefont{M.}~\bibnamefont{Xie}},
  \bibinfo{author}{\bibfnamefont{M.~A.} \bibnamefont{Aamir}},
  \bibinfo{author}{\bibfnamefont{I.}~\bibnamefont{Das}},
  \bibinfo{author}{\bibfnamefont{C.}~\bibnamefont{Urgell}},
  \bibinfo{author}{\bibfnamefont{K.}~\bibnamefont{Watanabe}},
  \bibinfo{author}{\bibfnamefont{T.}~\bibnamefont{Taniguchi}},
  \bibinfo{author}{\bibfnamefont{G.}~\bibnamefont{Zhang}},
  \bibnamefont{et~al.}, \bibinfo{journal}{Nature}
  \textbf{\bibinfo{volume}{574}}, \bibinfo{pages}{653} (\bibinfo{year}{2019}).

\bibitem[{\citenamefont{Stepanov
  et~al.}(2020{\natexlab{a}})\citenamefont{Stepanov, Das, Lu, Fahimniya,
  Watanabe, Taniguchi, Koppens, Lischner, Levitov, and
  Efetov}}]{efetov-nature20}
\bibinfo{author}{\bibfnamefont{P.}~\bibnamefont{Stepanov}},
  \bibinfo{author}{\bibfnamefont{I.}~\bibnamefont{Das}},
  \bibinfo{author}{\bibfnamefont{X.}~\bibnamefont{Lu}},
  \bibinfo{author}{\bibfnamefont{A.}~\bibnamefont{Fahimniya}},
  \bibinfo{author}{\bibfnamefont{K.}~\bibnamefont{Watanabe}},
  \bibinfo{author}{\bibfnamefont{T.}~\bibnamefont{Taniguchi}},
  \bibinfo{author}{\bibfnamefont{F.~H.~L.} \bibnamefont{Koppens}},
  \bibinfo{author}{\bibfnamefont{J.}~\bibnamefont{Lischner}},
  \bibinfo{author}{\bibfnamefont{L.}~\bibnamefont{Levitov}}, \bibnamefont{and}
  \bibinfo{author}{\bibfnamefont{D.~K.} \bibnamefont{Efetov}},
  \bibinfo{journal}{Nature} \textbf{\bibinfo{volume}{583}},
  \bibinfo{pages}{375} (\bibinfo{year}{2020}{\natexlab{a}}), ISSN
  \bibinfo{issn}{1476-4687}.

\bibitem[{\citenamefont{Saito et~al.}(2020)\citenamefont{Saito, Ge, Watanabe,
  Taniguchi, and Young}}]{young-tbg-np20}
\bibinfo{author}{\bibfnamefont{Y.}~\bibnamefont{Saito}},
  \bibinfo{author}{\bibfnamefont{J.}~\bibnamefont{Ge}},
  \bibinfo{author}{\bibfnamefont{K.}~\bibnamefont{Watanabe}},
  \bibinfo{author}{\bibfnamefont{T.}~\bibnamefont{Taniguchi}},
  \bibnamefont{and} \bibinfo{author}{\bibfnamefont{A.~F.} \bibnamefont{Young}},
  \bibinfo{journal}{Nature Physics} \textbf{\bibinfo{volume}{16}},
  \bibinfo{pages}{926} (\bibinfo{year}{2020}), ISSN \bibinfo{issn}{1745-2481}.

\bibitem[{\citenamefont{Liu et~al.}(2021)\citenamefont{Liu, Wang, Watanabe,
  Taniguchi, Vafek, and Li}}]{li-tbg-science21}
\bibinfo{author}{\bibfnamefont{X.}~\bibnamefont{Liu}},
  \bibinfo{author}{\bibfnamefont{Z.}~\bibnamefont{Wang}},
  \bibinfo{author}{\bibfnamefont{K.}~\bibnamefont{Watanabe}},
  \bibinfo{author}{\bibfnamefont{T.}~\bibnamefont{Taniguchi}},
  \bibinfo{author}{\bibfnamefont{O.}~\bibnamefont{Vafek}}, \bibnamefont{and}
  \bibinfo{author}{\bibfnamefont{J.}~\bibnamefont{Li}},
  \bibinfo{journal}{Science} \textbf{\bibinfo{volume}{371}},
  \bibinfo{pages}{1261} (\bibinfo{year}{2021}).

\bibitem[{\citenamefont{Cao et~al.}(2021)\citenamefont{Cao, Rodan-Legrain,
  Park, Yuan, Watanabe, Taniguchi, Fernandes, Fu, and
  Jarillo-Herrero}}]{cao-tbg-nematic-science21}
\bibinfo{author}{\bibfnamefont{Y.}~\bibnamefont{Cao}},
  \bibinfo{author}{\bibfnamefont{D.}~\bibnamefont{Rodan-Legrain}},
  \bibinfo{author}{\bibfnamefont{J.~M.} \bibnamefont{Park}},
  \bibinfo{author}{\bibfnamefont{N.~F.} \bibnamefont{Yuan}},
  \bibinfo{author}{\bibfnamefont{K.}~\bibnamefont{Watanabe}},
  \bibinfo{author}{\bibfnamefont{T.}~\bibnamefont{Taniguchi}},
  \bibinfo{author}{\bibfnamefont{R.~M.} \bibnamefont{Fernandes}},
  \bibinfo{author}{\bibfnamefont{L.}~\bibnamefont{Fu}}, \bibnamefont{and}
  \bibinfo{author}{\bibfnamefont{P.}~\bibnamefont{Jarillo-Herrero}},
  \bibinfo{journal}{science} \textbf{\bibinfo{volume}{372}},
  \bibinfo{pages}{264} (\bibinfo{year}{2021}).

\bibitem[{\citenamefont{Cao et~al.}(2018{\natexlab{b}})\citenamefont{Cao,
  Fatemi, Demir, Fang, Tomarken, Luo, Sanchez-Yamagishi, Watanabe, Taniguchi,
  Kaxiras et~al.}}]{cao-nature18-mott}
\bibinfo{author}{\bibfnamefont{Y.}~\bibnamefont{Cao}},
  \bibinfo{author}{\bibfnamefont{V.}~\bibnamefont{Fatemi}},
  \bibinfo{author}{\bibfnamefont{A.}~\bibnamefont{Demir}},
  \bibinfo{author}{\bibfnamefont{S.}~\bibnamefont{Fang}},
  \bibinfo{author}{\bibfnamefont{S.~L.} \bibnamefont{Tomarken}},
  \bibinfo{author}{\bibfnamefont{J.~Y.} \bibnamefont{Luo}},
  \bibinfo{author}{\bibfnamefont{J.~D.} \bibnamefont{Sanchez-Yamagishi}},
  \bibinfo{author}{\bibfnamefont{K.}~\bibnamefont{Watanabe}},
  \bibinfo{author}{\bibfnamefont{T.}~\bibnamefont{Taniguchi}},
  \bibinfo{author}{\bibfnamefont{E.}~\bibnamefont{Kaxiras}},
  \bibnamefont{et~al.}, \bibinfo{journal}{Nature}
  \textbf{\bibinfo{volume}{556}}, \bibinfo{pages}{80}
  (\bibinfo{year}{2018}{\natexlab{b}}).

\bibitem[{\citenamefont{Kerelsky et~al.}(2019)\citenamefont{Kerelsky, McGilly,
  Kennes, Xian, Yankowitz, Chen, Watanabe, Taniguchi, Hone, Dean
  et~al.}}]{tbg-stm-pasupathy19}
\bibinfo{author}{\bibfnamefont{A.}~\bibnamefont{Kerelsky}},
  \bibinfo{author}{\bibfnamefont{L.~J.} \bibnamefont{McGilly}},
  \bibinfo{author}{\bibfnamefont{D.~M.} \bibnamefont{Kennes}},
  \bibinfo{author}{\bibfnamefont{L.}~\bibnamefont{Xian}},
  \bibinfo{author}{\bibfnamefont{M.}~\bibnamefont{Yankowitz}},
  \bibinfo{author}{\bibfnamefont{S.}~\bibnamefont{Chen}},
  \bibinfo{author}{\bibfnamefont{K.}~\bibnamefont{Watanabe}},
  \bibinfo{author}{\bibfnamefont{T.}~\bibnamefont{Taniguchi}},
  \bibinfo{author}{\bibfnamefont{J.}~\bibnamefont{Hone}},
  \bibinfo{author}{\bibfnamefont{C.}~\bibnamefont{Dean}}, \bibnamefont{et~al.},
  \bibinfo{journal}{Nature} \textbf{\bibinfo{volume}{572}}, \bibinfo{pages}{95}
  (\bibinfo{year}{2019}).

\bibitem[{\citenamefont{{Jiang} et~al.}(2019)\citenamefont{{Jiang}, {Lai},
  {Watanabe}, {Taniguchi}, {Haule}, {Mao}, and {Andrei}}}]{tbg-stm-andrei19}
\bibinfo{author}{\bibfnamefont{Y.}~\bibnamefont{{Jiang}}},
  \bibinfo{author}{\bibfnamefont{X.}~\bibnamefont{{Lai}}},
  \bibinfo{author}{\bibfnamefont{K.}~\bibnamefont{{Watanabe}}},
  \bibinfo{author}{\bibfnamefont{T.}~\bibnamefont{{Taniguchi}}},
  \bibinfo{author}{\bibfnamefont{K.}~\bibnamefont{{Haule}}},
  \bibinfo{author}{\bibfnamefont{J.}~\bibnamefont{{Mao}}}, \bibnamefont{and}
  \bibinfo{author}{\bibfnamefont{E.~Y.} \bibnamefont{{Andrei}}},
  \bibinfo{journal}{\nat} \textbf{\bibinfo{volume}{573}}, \bibinfo{pages}{91}
  (\bibinfo{year}{2019}).

\bibitem[{\citenamefont{{Xie} et~al.}(2019)\citenamefont{{Xie}, {Lian},
  {J{\"a}ck}, {Liu}, {Chiu}, {Watanabe}, {Taniguchi}, {Bernevig}, and
  {Yazdani}}}]{tbg-stm-yazdani19}
\bibinfo{author}{\bibfnamefont{Y.}~\bibnamefont{{Xie}}},
  \bibinfo{author}{\bibfnamefont{B.}~\bibnamefont{{Lian}}},
  \bibinfo{author}{\bibfnamefont{B.}~\bibnamefont{{J{\"a}ck}}},
  \bibinfo{author}{\bibfnamefont{X.}~\bibnamefont{{Liu}}},
  \bibinfo{author}{\bibfnamefont{C.-L.} \bibnamefont{{Chiu}}},
  \bibinfo{author}{\bibfnamefont{K.}~\bibnamefont{{Watanabe}}},
  \bibinfo{author}{\bibfnamefont{T.}~\bibnamefont{{Taniguchi}}},
  \bibinfo{author}{\bibfnamefont{B.~A.} \bibnamefont{{Bernevig}}},
  \bibnamefont{and}
  \bibinfo{author}{\bibfnamefont{A.}~\bibnamefont{{Yazdani}}},
  \bibinfo{journal}{\nat} \textbf{\bibinfo{volume}{572}}, \bibinfo{pages}{101}
  (\bibinfo{year}{2019}).

\bibitem[{\citenamefont{Choi et~al.}(2019)\citenamefont{Choi, Kemmer, Peng,
  Thomson, Arora, Polski, Zhang, Ren, Alicea, Refael
  et~al.}}]{tbg-stm-caltech19}
\bibinfo{author}{\bibfnamefont{Y.}~\bibnamefont{Choi}},
  \bibinfo{author}{\bibfnamefont{J.}~\bibnamefont{Kemmer}},
  \bibinfo{author}{\bibfnamefont{Y.}~\bibnamefont{Peng}},
  \bibinfo{author}{\bibfnamefont{A.}~\bibnamefont{Thomson}},
  \bibinfo{author}{\bibfnamefont{H.}~\bibnamefont{Arora}},
  \bibinfo{author}{\bibfnamefont{R.}~\bibnamefont{Polski}},
  \bibinfo{author}{\bibfnamefont{Y.}~\bibnamefont{Zhang}},
  \bibinfo{author}{\bibfnamefont{H.}~\bibnamefont{Ren}},
  \bibinfo{author}{\bibfnamefont{J.}~\bibnamefont{Alicea}},
  \bibinfo{author}{\bibfnamefont{G.}~\bibnamefont{Refael}},
  \bibnamefont{et~al.}, \bibinfo{journal}{Nature Physics} pp.
  \bibinfo{pages}{1--7} (\bibinfo{year}{2019}).

\bibitem[{\citenamefont{Serlin et~al.}(2019)\citenamefont{Serlin, Tschirhart,
  Polshyn, Zhang, Zhu, Watanabe, Taniguchi, Balents, and
  Young}}]{young-tbg-science19}
\bibinfo{author}{\bibfnamefont{M.}~\bibnamefont{Serlin}},
  \bibinfo{author}{\bibfnamefont{C.}~\bibnamefont{Tschirhart}},
  \bibinfo{author}{\bibfnamefont{H.}~\bibnamefont{Polshyn}},
  \bibinfo{author}{\bibfnamefont{Y.}~\bibnamefont{Zhang}},
  \bibinfo{author}{\bibfnamefont{J.}~\bibnamefont{Zhu}},
  \bibinfo{author}{\bibfnamefont{K.}~\bibnamefont{Watanabe}},
  \bibinfo{author}{\bibfnamefont{T.}~\bibnamefont{Taniguchi}},
  \bibinfo{author}{\bibfnamefont{L.}~\bibnamefont{Balents}}, \bibnamefont{and}
  \bibinfo{author}{\bibfnamefont{A.}~\bibnamefont{Young}},
  \bibinfo{journal}{Science}  (\bibinfo{year}{2019}).

\bibitem[{\citenamefont{{Sharpe} et~al.}(2019)\citenamefont{{Sharpe}, {Fox},
  {Barnard}, {Finney}, {Watanabe}, {Taniguchi}, {Kastner}, and
  {Goldhaber-Gordon}}}]{sharpe-science-19}
\bibinfo{author}{\bibfnamefont{A.~L.} \bibnamefont{{Sharpe}}},
  \bibinfo{author}{\bibfnamefont{E.~J.} \bibnamefont{{Fox}}},
  \bibinfo{author}{\bibfnamefont{A.~W.} \bibnamefont{{Barnard}}},
  \bibinfo{author}{\bibfnamefont{J.}~\bibnamefont{{Finney}}},
  \bibinfo{author}{\bibfnamefont{K.}~\bibnamefont{{Watanabe}}},
  \bibinfo{author}{\bibfnamefont{T.}~\bibnamefont{{Taniguchi}}},
  \bibinfo{author}{\bibfnamefont{M.~A.} \bibnamefont{{Kastner}}},
  \bibnamefont{and}
  \bibinfo{author}{\bibfnamefont{D.}~\bibnamefont{{Goldhaber-Gordon}}},
  \bibinfo{journal}{Science} \textbf{\bibinfo{volume}{365}},
  \bibinfo{pages}{605} (\bibinfo{year}{2019}).

\bibitem[{\citenamefont{Stepanov
  et~al.}(2020{\natexlab{b}})\citenamefont{Stepanov, Xie, Taniguchi, Watanabe,
  Lu, MacDonald, Bernevig, and Efetov}}]{efetov-arxiv20}
\bibinfo{author}{\bibfnamefont{P.}~\bibnamefont{Stepanov}},
  \bibinfo{author}{\bibfnamefont{M.}~\bibnamefont{Xie}},
  \bibinfo{author}{\bibfnamefont{T.}~\bibnamefont{Taniguchi}},
  \bibinfo{author}{\bibfnamefont{K.}~\bibnamefont{Watanabe}},
  \bibinfo{author}{\bibfnamefont{X.}~\bibnamefont{Lu}},
  \bibinfo{author}{\bibfnamefont{A.~H.} \bibnamefont{MacDonald}},
  \bibinfo{author}{\bibfnamefont{B.~A.} \bibnamefont{Bernevig}},
  \bibnamefont{and} \bibinfo{author}{\bibfnamefont{D.~K.} \bibnamefont{Efetov}}
  (\bibinfo{year}{2020}{\natexlab{b}}), \eprint{2012.15126}.

\bibitem[{\citenamefont{Polshyn et~al.}(2020)\citenamefont{Polshyn, Zhu, Kumar,
  Zhang, Yang, Tschirhart, Serlin, Watanabe, Taniguchi, MacDonald
  et~al.}}]{young-monobi-nature20}
\bibinfo{author}{\bibfnamefont{H.}~\bibnamefont{Polshyn}},
  \bibinfo{author}{\bibfnamefont{J.}~\bibnamefont{Zhu}},
  \bibinfo{author}{\bibfnamefont{M.}~\bibnamefont{Kumar}},
  \bibinfo{author}{\bibfnamefont{Y.}~\bibnamefont{Zhang}},
  \bibinfo{author}{\bibfnamefont{F.}~\bibnamefont{Yang}},
  \bibinfo{author}{\bibfnamefont{C.}~\bibnamefont{Tschirhart}},
  \bibinfo{author}{\bibfnamefont{M.}~\bibnamefont{Serlin}},
  \bibinfo{author}{\bibfnamefont{K.}~\bibnamefont{Watanabe}},
  \bibinfo{author}{\bibfnamefont{T.}~\bibnamefont{Taniguchi}},
  \bibinfo{author}{\bibfnamefont{A.}~\bibnamefont{MacDonald}},
  \bibnamefont{et~al.}, \bibinfo{journal}{Nature} pp. \bibinfo{pages}{1--5}
  (\bibinfo{year}{2020}).

\bibitem[{\citenamefont{Chen et~al.}(2020)\citenamefont{Chen, He, Zhang, Hsieh,
  Fei, Watanabe, Taniguchi, Cobden, Xu, Dean et~al.}}]{Yankowitz-monobi-np2020}
\bibinfo{author}{\bibfnamefont{S.}~\bibnamefont{Chen}},
  \bibinfo{author}{\bibfnamefont{M.}~\bibnamefont{He}},
  \bibinfo{author}{\bibfnamefont{Y.-H.} \bibnamefont{Zhang}},
  \bibinfo{author}{\bibfnamefont{V.}~\bibnamefont{Hsieh}},
  \bibinfo{author}{\bibfnamefont{Z.}~\bibnamefont{Fei}},
  \bibinfo{author}{\bibfnamefont{K.}~\bibnamefont{Watanabe}},
  \bibinfo{author}{\bibfnamefont{T.}~\bibnamefont{Taniguchi}},
  \bibinfo{author}{\bibfnamefont{D.~H.} \bibnamefont{Cobden}},
  \bibinfo{author}{\bibfnamefont{X.}~\bibnamefont{Xu}},
  \bibinfo{author}{\bibfnamefont{C.~R.} \bibnamefont{Dean}},
  \bibnamefont{et~al.}, \bibinfo{journal}{Nat. Phys.}  (\bibinfo{year}{2020}).

\bibitem[{\citenamefont{Xu et~al.}(2021)\citenamefont{Xu, Al~Ezzi,
  Balakrishnan, Garcia-Ruiz, Tsim, Mullan, Barrier, Xin, Piot, Taniguchi
  et~al.}}]{shi-tbmg-np21}
\bibinfo{author}{\bibfnamefont{S.}~\bibnamefont{Xu}},
  \bibinfo{author}{\bibfnamefont{M.~M.} \bibnamefont{Al~Ezzi}},
  \bibinfo{author}{\bibfnamefont{N.}~\bibnamefont{Balakrishnan}},
  \bibinfo{author}{\bibfnamefont{A.}~\bibnamefont{Garcia-Ruiz}},
  \bibinfo{author}{\bibfnamefont{B.}~\bibnamefont{Tsim}},
  \bibinfo{author}{\bibfnamefont{C.}~\bibnamefont{Mullan}},
  \bibinfo{author}{\bibfnamefont{J.}~\bibnamefont{Barrier}},
  \bibinfo{author}{\bibfnamefont{N.}~\bibnamefont{Xin}},
  \bibinfo{author}{\bibfnamefont{B.~A.} \bibnamefont{Piot}},
  \bibinfo{author}{\bibfnamefont{T.}~\bibnamefont{Taniguchi}},
  \bibnamefont{et~al.}, \bibinfo{journal}{Nature Physics}
  (\bibinfo{year}{2021}), ISSN \bibinfo{issn}{1745-2481}.

\bibitem[{\citenamefont{Liu et~al.}(2020)\citenamefont{Liu, Hao, Khalaf, Lee,
  Ronen, Yoo, Haei~Najafabadi, Watanabe, Taniguchi, Vishwanath
  et~al.}}]{kim-tdbg-nature20}
\bibinfo{author}{\bibfnamefont{X.}~\bibnamefont{Liu}},
  \bibinfo{author}{\bibfnamefont{Z.}~\bibnamefont{Hao}},
  \bibinfo{author}{\bibfnamefont{E.}~\bibnamefont{Khalaf}},
  \bibinfo{author}{\bibfnamefont{J.~Y.} \bibnamefont{Lee}},
  \bibinfo{author}{\bibfnamefont{Y.}~\bibnamefont{Ronen}},
  \bibinfo{author}{\bibfnamefont{H.}~\bibnamefont{Yoo}},
  \bibinfo{author}{\bibfnamefont{D.}~\bibnamefont{Haei~Najafabadi}},
  \bibinfo{author}{\bibfnamefont{K.}~\bibnamefont{Watanabe}},
  \bibinfo{author}{\bibfnamefont{T.}~\bibnamefont{Taniguchi}},
  \bibinfo{author}{\bibfnamefont{A.}~\bibnamefont{Vishwanath}},
  \bibnamefont{et~al.}, \bibinfo{journal}{Nature}
  \textbf{\bibinfo{volume}{583}}, \bibinfo{pages}{221} (\bibinfo{year}{2020}).

\bibitem[{\citenamefont{Shen et~al.}(2020)\citenamefont{Shen, Chu, Wu, Li,
  Wang, Zhao, Tang, Liu, Tian, Watanabe et~al.}}]{zhang-tdbg-np20}
\bibinfo{author}{\bibfnamefont{C.}~\bibnamefont{Shen}},
  \bibinfo{author}{\bibfnamefont{Y.}~\bibnamefont{Chu}},
  \bibinfo{author}{\bibfnamefont{Q.}~\bibnamefont{Wu}},
  \bibinfo{author}{\bibfnamefont{N.}~\bibnamefont{Li}},
  \bibinfo{author}{\bibfnamefont{S.}~\bibnamefont{Wang}},
  \bibinfo{author}{\bibfnamefont{Y.}~\bibnamefont{Zhao}},
  \bibinfo{author}{\bibfnamefont{J.}~\bibnamefont{Tang}},
  \bibinfo{author}{\bibfnamefont{J.}~\bibnamefont{Liu}},
  \bibinfo{author}{\bibfnamefont{J.}~\bibnamefont{Tian}},
  \bibinfo{author}{\bibfnamefont{K.}~\bibnamefont{Watanabe}},
  \bibnamefont{et~al.}, \bibinfo{journal}{Nature Physics}
  \textbf{\bibinfo{volume}{16}}, \bibinfo{pages}{520} (\bibinfo{year}{2020}),
  ISSN \bibinfo{issn}{1745-2481}.

\bibitem[{\citenamefont{Cao et~al.}(2020)\citenamefont{Cao, Rodan-Legrain,
  Rubies-Bigorda, Park, Watanabe, Taniguchi, and
  Jarillo-Herrero}}]{cao-tdbg-nature20}
\bibinfo{author}{\bibfnamefont{Y.}~\bibnamefont{Cao}},
  \bibinfo{author}{\bibfnamefont{D.}~\bibnamefont{Rodan-Legrain}},
  \bibinfo{author}{\bibfnamefont{O.}~\bibnamefont{Rubies-Bigorda}},
  \bibinfo{author}{\bibfnamefont{J.~M.} \bibnamefont{Park}},
  \bibinfo{author}{\bibfnamefont{K.}~\bibnamefont{Watanabe}},
  \bibinfo{author}{\bibfnamefont{T.}~\bibnamefont{Taniguchi}},
  \bibnamefont{and}
  \bibinfo{author}{\bibfnamefont{P.}~\bibnamefont{Jarillo-Herrero}},
  \bibinfo{journal}{Nature}  (\bibinfo{year}{2020}), ISSN
  \bibinfo{issn}{1476-4687}.

\bibitem[{\citenamefont{Rubio-Verd{\'u}
  et~al.}(2020)\citenamefont{Rubio-Verd{\'u}, Turkel, Song, Klebl, Samajdar,
  Scheurer, Venderbos, Watanabe, Taniguchi, Ochoa
  et~al.}}]{Pasupathy-nematic-tdbg-arxiv20}
\bibinfo{author}{\bibfnamefont{C.}~\bibnamefont{Rubio-Verd{\'u}}},
  \bibinfo{author}{\bibfnamefont{S.}~\bibnamefont{Turkel}},
  \bibinfo{author}{\bibfnamefont{L.}~\bibnamefont{Song}},
  \bibinfo{author}{\bibfnamefont{L.}~\bibnamefont{Klebl}},
  \bibinfo{author}{\bibfnamefont{R.}~\bibnamefont{Samajdar}},
  \bibinfo{author}{\bibfnamefont{M.~S.} \bibnamefont{Scheurer}},
  \bibinfo{author}{\bibfnamefont{J.~W.~F.} \bibnamefont{Venderbos}},
  \bibinfo{author}{\bibfnamefont{K.}~\bibnamefont{Watanabe}},
  \bibinfo{author}{\bibfnamefont{T.}~\bibnamefont{Taniguchi}},
  \bibinfo{author}{\bibfnamefont{H.}~\bibnamefont{Ochoa}},
  \bibnamefont{et~al.}, \bibinfo{journal}{arXiv preprint arXiv:2009.11645}
  (\bibinfo{year}{2020}).

\bibitem[{\citenamefont{Chen et~al.}(2019{\natexlab{a}})\citenamefont{Chen,
  Sharpe, Gallagher, Rosen, Fox, Jiang, Lyu, Li, Watanabe, Taniguchi
  et~al.}}]{chen-hbn-trilayer-nature19}
\bibinfo{author}{\bibfnamefont{G.}~\bibnamefont{Chen}},
  \bibinfo{author}{\bibfnamefont{A.~L.} \bibnamefont{Sharpe}},
  \bibinfo{author}{\bibfnamefont{P.}~\bibnamefont{Gallagher}},
  \bibinfo{author}{\bibfnamefont{I.~T.} \bibnamefont{Rosen}},
  \bibinfo{author}{\bibfnamefont{E.~J.} \bibnamefont{Fox}},
  \bibinfo{author}{\bibfnamefont{L.}~\bibnamefont{Jiang}},
  \bibinfo{author}{\bibfnamefont{B.}~\bibnamefont{Lyu}},
  \bibinfo{author}{\bibfnamefont{H.}~\bibnamefont{Li}},
  \bibinfo{author}{\bibfnamefont{K.}~\bibnamefont{Watanabe}},
  \bibinfo{author}{\bibfnamefont{T.}~\bibnamefont{Taniguchi}},
  \bibnamefont{et~al.}, \bibinfo{journal}{Nature}
  \textbf{\bibinfo{volume}{572}}, \bibinfo{pages}{215}
  (\bibinfo{year}{2019}{\natexlab{a}}), ISSN \bibinfo{issn}{1476-4687}.

\bibitem[{\citenamefont{Chen et~al.}(2019{\natexlab{b}})\citenamefont{Chen,
  Jiang, Wu, Lyu, Li, Chittari, Watanabe, Taniguchi, Shi, Jung
  et~al.}}]{chen-trilayer-hbn-mott-np19}
\bibinfo{author}{\bibfnamefont{G.}~\bibnamefont{Chen}},
  \bibinfo{author}{\bibfnamefont{L.}~\bibnamefont{Jiang}},
  \bibinfo{author}{\bibfnamefont{S.}~\bibnamefont{Wu}},
  \bibinfo{author}{\bibfnamefont{B.}~\bibnamefont{Lyu}},
  \bibinfo{author}{\bibfnamefont{H.}~\bibnamefont{Li}},
  \bibinfo{author}{\bibfnamefont{B.~L.} \bibnamefont{Chittari}},
  \bibinfo{author}{\bibfnamefont{K.}~\bibnamefont{Watanabe}},
  \bibinfo{author}{\bibfnamefont{T.}~\bibnamefont{Taniguchi}},
  \bibinfo{author}{\bibfnamefont{Z.}~\bibnamefont{Shi}},
  \bibinfo{author}{\bibfnamefont{J.}~\bibnamefont{Jung}}, \bibnamefont{et~al.},
  \bibinfo{journal}{Nat. Phys.} \textbf{\bibinfo{volume}{15}},
  \bibinfo{pages}{237} (\bibinfo{year}{2019}{\natexlab{b}}).

\bibitem[{\citenamefont{Po et~al.}(2019)\citenamefont{Po, Zou, Senthil, and
  Vishwanath}}]{po-tbg-prb19}
\bibinfo{author}{\bibfnamefont{H.~C.} \bibnamefont{Po}},
  \bibinfo{author}{\bibfnamefont{L.}~\bibnamefont{Zou}},
  \bibinfo{author}{\bibfnamefont{T.}~\bibnamefont{Senthil}}, \bibnamefont{and}
  \bibinfo{author}{\bibfnamefont{A.}~\bibnamefont{Vishwanath}},
  \bibinfo{journal}{Phys. Rev. B} \textbf{\bibinfo{volume}{99}},
  \bibinfo{pages}{195455} (\bibinfo{year}{2019}).

\bibitem[{\citenamefont{Liu et~al.}(2019{\natexlab{a}})\citenamefont{Liu, Liu,
  and Dai}}]{jpliu-prb19}
\bibinfo{author}{\bibfnamefont{J.}~\bibnamefont{Liu}},
  \bibinfo{author}{\bibfnamefont{J.}~\bibnamefont{Liu}}, \bibnamefont{and}
  \bibinfo{author}{\bibfnamefont{X.}~\bibnamefont{Dai}},
  \bibinfo{journal}{Phys. Rev. B} \textbf{\bibinfo{volume}{99}},
  \bibinfo{pages}{155415} (\bibinfo{year}{2019}{\natexlab{a}}).

\bibitem[{\citenamefont{Bultinck et~al.}(2020)\citenamefont{Bultinck,
  Chatterjee, and Zaletel}}]{zaletel-tbg-2019}
\bibinfo{author}{\bibfnamefont{N.}~\bibnamefont{Bultinck}},
  \bibinfo{author}{\bibfnamefont{S.}~\bibnamefont{Chatterjee}},
  \bibnamefont{and} \bibinfo{author}{\bibfnamefont{M.~P.}
  \bibnamefont{Zaletel}}, \bibinfo{journal}{Phys. Rev. Lett.}
  \textbf{\bibinfo{volume}{124}}, \bibinfo{pages}{166601}
  (\bibinfo{year}{2020}).

\bibitem[{\citenamefont{Song et~al.}(2019)\citenamefont{Song, Wang, Shi, Li,
  Fang, and Bernevig}}]{song-tbg-prl19}
\bibinfo{author}{\bibfnamefont{Z.}~\bibnamefont{Song}},
  \bibinfo{author}{\bibfnamefont{Z.}~\bibnamefont{Wang}},
  \bibinfo{author}{\bibfnamefont{W.}~\bibnamefont{Shi}},
  \bibinfo{author}{\bibfnamefont{G.}~\bibnamefont{Li}},
  \bibinfo{author}{\bibfnamefont{C.}~\bibnamefont{Fang}}, \bibnamefont{and}
  \bibinfo{author}{\bibfnamefont{B.~A.} \bibnamefont{Bernevig}},
  \bibinfo{journal}{Phys. Rev. Lett.} \textbf{\bibinfo{volume}{123}},
  \bibinfo{pages}{036401} (\bibinfo{year}{2019}).

\bibitem[{\citenamefont{Ahn et~al.}(2019)\citenamefont{Ahn, Park, and
  Yang}}]{yang-tbg-prx19}
\bibinfo{author}{\bibfnamefont{J.}~\bibnamefont{Ahn}},
  \bibinfo{author}{\bibfnamefont{S.}~\bibnamefont{Park}}, \bibnamefont{and}
  \bibinfo{author}{\bibfnamefont{B.-J.} \bibnamefont{Yang}},
  \bibinfo{journal}{Phys. Rev. X} \textbf{\bibinfo{volume}{9}},
  \bibinfo{pages}{021013} (\bibinfo{year}{2019}).

\bibitem[{\citenamefont{Zhang et~al.}(2019{\natexlab{a}})\citenamefont{Zhang,
  Mao, and Senthil}}]{senthil-tbg-prr19}
\bibinfo{author}{\bibfnamefont{Y.-H.} \bibnamefont{Zhang}},
  \bibinfo{author}{\bibfnamefont{D.}~\bibnamefont{Mao}}, \bibnamefont{and}
  \bibinfo{author}{\bibfnamefont{T.}~\bibnamefont{Senthil}},
  \bibinfo{journal}{Phys. Rev. Research} \textbf{\bibinfo{volume}{1}},
  \bibinfo{pages}{033126} (\bibinfo{year}{2019}{\natexlab{a}}).

\bibitem[{\citenamefont{Liu et~al.}(2019{\natexlab{b}})\citenamefont{Liu, Ma,
  Gao, and Dai}}]{jpliu-prx19}
\bibinfo{author}{\bibfnamefont{J.}~\bibnamefont{Liu}},
  \bibinfo{author}{\bibfnamefont{Z.}~\bibnamefont{Ma}},
  \bibinfo{author}{\bibfnamefont{J.}~\bibnamefont{Gao}}, \bibnamefont{and}
  \bibinfo{author}{\bibfnamefont{X.}~\bibnamefont{Dai}},
  \bibinfo{journal}{Phys. Rev. X} \textbf{\bibinfo{volume}{9}},
  \bibinfo{pages}{031021} (\bibinfo{year}{2019}{\natexlab{b}}).

\bibitem[{\citenamefont{Bistritzer and MacDonald}(2011)}]{macdonald-pnas11}
\bibinfo{author}{\bibfnamefont{R.}~\bibnamefont{Bistritzer}} \bibnamefont{and}
  \bibinfo{author}{\bibfnamefont{A.~H.} \bibnamefont{MacDonald}},
  \bibinfo{journal}{Proceedings of the National Academy of Sciences}
  \textbf{\bibinfo{volume}{108}}, \bibinfo{pages}{12233}
  (\bibinfo{year}{2011}).

\bibitem[{\citenamefont{Lopes~dos Santos et~al.}(2012)\citenamefont{Lopes~dos
  Santos, Peres, and Castro~Neto}}]{castro-neto-prb12}
\bibinfo{author}{\bibfnamefont{J.~M.~B.} \bibnamefont{Lopes~dos Santos}},
  \bibinfo{author}{\bibfnamefont{N.~M.~R.} \bibnamefont{Peres}},
  \bibnamefont{and} \bibinfo{author}{\bibfnamefont{A.~H.}
  \bibnamefont{Castro~Neto}}, \bibinfo{journal}{Phys. Rev. B}
  \textbf{\bibinfo{volume}{86}}, \bibinfo{pages}{155449}
  (\bibinfo{year}{2012}).

\bibitem[{\citenamefont{Po et~al.}(2018)\citenamefont{Po, Zou, Vishwanath, and
  Senthil}}]{po-prx18}
\bibinfo{author}{\bibfnamefont{H.~C.} \bibnamefont{Po}},
  \bibinfo{author}{\bibfnamefont{L.}~\bibnamefont{Zou}},
  \bibinfo{author}{\bibfnamefont{A.}~\bibnamefont{Vishwanath}},
  \bibnamefont{and} \bibinfo{author}{\bibfnamefont{T.}~\bibnamefont{Senthil}},
  \bibinfo{journal}{Phys. Rev. X} \textbf{\bibinfo{volume}{8}},
  \bibinfo{pages}{031089} (\bibinfo{year}{2018}).

\bibitem[{\citenamefont{Koshino}(2019)}]{koshino-tdbg-prb19}
\bibinfo{author}{\bibfnamefont{M.}~\bibnamefont{Koshino}},
  \bibinfo{journal}{Phys. Rev. B} \textbf{\bibinfo{volume}{99}},
  \bibinfo{pages}{235406} (\bibinfo{year}{2019}).

\bibitem[{\citenamefont{King-Smith and Vanderbilt}(1993)}]{kingsmith-prb93}
\bibinfo{author}{\bibfnamefont{R.}~\bibnamefont{King-Smith}} \bibnamefont{and}
  \bibinfo{author}{\bibfnamefont{D.}~\bibnamefont{Vanderbilt}},
  \bibinfo{journal}{Phys. Rev. B} \textbf{\bibinfo{volume}{47}},
  \bibinfo{pages}{1651} (\bibinfo{year}{1993}).

\bibitem[{\citenamefont{Resta and Vanderbilt}(2007)}]{resta07}
\bibinfo{author}{\bibfnamefont{R.}~\bibnamefont{Resta}} \bibnamefont{and}
  \bibinfo{author}{\bibfnamefont{D.}~\bibnamefont{Vanderbilt}}, in
  \emph{\bibinfo{booktitle}{Physics of Ferroelectrics: a Modern Perspective}},
  edited by \bibinfo{editor}{\bibfnamefont{K.~M.} \bibnamefont{Rabe}},
  \bibinfo{editor}{\bibfnamefont{C.~H.} \bibnamefont{Ahn}}, \bibnamefont{and}
  \bibinfo{editor}{\bibfnamefont{J.-M.} \bibnamefont{Triscone}}
  (\bibinfo{publisher}{Springer-Verlag}, \bibinfo{address}{Berlin},
  \bibinfo{year}{2007}).

\bibitem[{\citenamefont{Coh and Vanderbilt}(2009)}]{polarization-chern}
\bibinfo{author}{\bibfnamefont{S.}~\bibnamefont{Coh}} \bibnamefont{and}
  \bibinfo{author}{\bibfnamefont{D.}~\bibnamefont{Vanderbilt}},
  \bibinfo{journal}{Phys. Rev. Lett.} \textbf{\bibinfo{volume}{102}},
  \bibinfo{pages}{107603} (\bibinfo{year}{2009}).

\bibitem[{\citenamefont{Vanderbilt}(2000)}]{vanderbilt-piezo-00}
\bibinfo{author}{\bibfnamefont{D.}~\bibnamefont{Vanderbilt}},
  \bibinfo{journal}{Journal of Physics and Chemistry of Solids}
  \textbf{\bibinfo{volume}{61}}, \bibinfo{pages}{147} (\bibinfo{year}{2000}),
  ISSN \bibinfo{issn}{0022-3697}.

\bibitem[{\citenamefont{Bi et~al.}(2019)\citenamefont{Bi, Yuan, and
  Fu}}]{bi-tbg-strain}
\bibinfo{author}{\bibfnamefont{Z.}~\bibnamefont{Bi}},
  \bibinfo{author}{\bibfnamefont{N.~F.~Q.} \bibnamefont{Yuan}},
  \bibnamefont{and} \bibinfo{author}{\bibfnamefont{L.}~\bibnamefont{Fu}},
  \bibinfo{journal}{Phys. Rev. B} \textbf{\bibinfo{volume}{100}},
  \bibinfo{pages}{035448} (\bibinfo{year}{2019}).

\bibitem[{\citenamefont{Yu and Liu}(2020)}]{piezo_TQPT_Natcomm}
\bibinfo{author}{\bibfnamefont{J.}~\bibnamefont{Yu}} \bibnamefont{and}
  \bibinfo{author}{\bibfnamefont{C.}~\bibnamefont{Liu}},
  \bibinfo{journal}{Nature communications} \textbf{\bibinfo{volume}{11}},
  \bibinfo{pages}{2290} (\bibinfo{year}{2020}).

\bibitem[{\citenamefont{Zhang et~al.}(2019{\natexlab{b}})\citenamefont{Zhang,
  Mao, and Senthil}}]{senthil-hbn-tbg-prr19}
\bibinfo{author}{\bibfnamefont{Y.-H.} \bibnamefont{Zhang}},
  \bibinfo{author}{\bibfnamefont{D.}~\bibnamefont{Mao}}, \bibnamefont{and}
  \bibinfo{author}{\bibfnamefont{T.}~\bibnamefont{Senthil}},
  \bibinfo{journal}{Phys. Rev. Research} \textbf{\bibinfo{volume}{1}},
  \bibinfo{pages}{033126} (\bibinfo{year}{2019}{\natexlab{b}}).

\bibitem[{\citenamefont{Moon and Koshino}(2014)}]{koshino-prb14}
\bibinfo{author}{\bibfnamefont{P.}~\bibnamefont{Moon}} \bibnamefont{and}
  \bibinfo{author}{\bibfnamefont{M.}~\bibnamefont{Koshino}},
  \bibinfo{journal}{Phys. Rev. B} \textbf{\bibinfo{volume}{90}},
  \bibinfo{pages}{155406} (\bibinfo{year}{2014}).

\bibitem[{\citenamefont{Jung et~al.}(2014)\citenamefont{Jung, Raoux, Qiao, and
  MacDonald}}]{jung-prb14}
\bibinfo{author}{\bibfnamefont{J.}~\bibnamefont{Jung}},
  \bibinfo{author}{\bibfnamefont{A.}~\bibnamefont{Raoux}},
  \bibinfo{author}{\bibfnamefont{Z.}~\bibnamefont{Qiao}}, \bibnamefont{and}
  \bibinfo{author}{\bibfnamefont{A.~H.} \bibnamefont{MacDonald}},
  \bibinfo{journal}{Phys. Rev. B} \textbf{\bibinfo{volume}{89}},
  \bibinfo{pages}{205414} (\bibinfo{year}{2014}).

\bibitem[{sup()}]{supp_info}
\bibinfo{note}{See Appendix for: (a) the detailed presentation of the continuum
  models for the twisted bilayer, and twisted multilayer graphene systems; (b)
  the results of piezoelectric response from atomistic tight-binding
  calculations; and (c) the derivations for the strain induced vector fields in
  graphene;}.

\bibitem[{\citenamefont{He et~al.}(2020)\citenamefont{He, Li, Cai, Liu,
  Watanabe, Taniguchi, Xu, and Yankowitz}}]{Yankowitz-doublebi-np2020}
\bibinfo{author}{\bibfnamefont{M.}~\bibnamefont{He}},
  \bibinfo{author}{\bibfnamefont{Y.}~\bibnamefont{Li}},
  \bibinfo{author}{\bibfnamefont{J.}~\bibnamefont{Cai}},
  \bibinfo{author}{\bibfnamefont{Y.}~\bibnamefont{Liu}},
  \bibinfo{author}{\bibfnamefont{K.}~\bibnamefont{Watanabe}},
  \bibinfo{author}{\bibfnamefont{T.}~\bibnamefont{Taniguchi}},
  \bibinfo{author}{\bibfnamefont{X.}~\bibnamefont{Xu}}, \bibnamefont{and}
  \bibinfo{author}{\bibfnamefont{M.}~\bibnamefont{Yankowitz}},
  \bibinfo{journal}{Nat. Phys.}  (\bibinfo{year}{2020}).

\bibitem[{\citenamefont{Moon and Koshino}(2013)}]{moon-tbg-prb13}
\bibinfo{author}{\bibfnamefont{P.}~\bibnamefont{Moon}} \bibnamefont{and}
  \bibinfo{author}{\bibfnamefont{M.}~\bibnamefont{Koshino}},
  \bibinfo{journal}{Physical Review B} \textbf{\bibinfo{volume}{87}},
  \bibinfo{pages}{205404} (\bibinfo{year}{2013}).

\bibitem[{\citenamefont{Marzari and Vanderbilt}(1997)}]{MLWF-1}
\bibinfo{author}{\bibfnamefont{N.}~\bibnamefont{Marzari}} \bibnamefont{and}
  \bibinfo{author}{\bibfnamefont{D.}~\bibnamefont{Vanderbilt}},
  \bibinfo{journal}{Phys. Rev. B} \textbf{\bibinfo{volume}{56}},
  \bibinfo{pages}{12847} (\bibinfo{year}{1997}).

\bibitem[{\citenamefont{Koshino et~al.}(2018)\citenamefont{Koshino, Yuan,
  Koretsune, Ochi, Kuroki, and Fu}}]{koshino-prx18}
\bibinfo{author}{\bibfnamefont{M.}~\bibnamefont{Koshino}},
  \bibinfo{author}{\bibfnamefont{N.~F.~Q.} \bibnamefont{Yuan}},
  \bibinfo{author}{\bibfnamefont{T.}~\bibnamefont{Koretsune}},
  \bibinfo{author}{\bibfnamefont{M.}~\bibnamefont{Ochi}},
  \bibinfo{author}{\bibfnamefont{K.}~\bibnamefont{Kuroki}}, \bibnamefont{and}
  \bibinfo{author}{\bibfnamefont{L.}~\bibnamefont{Fu}}, \bibinfo{journal}{Phys.
  Rev. X} \textbf{\bibinfo{volume}{8}}, \bibinfo{pages}{031087}
  (\bibinfo{year}{2018}).

\bibitem[{\citenamefont{Lee et~al.}(2008)\citenamefont{Lee, Lee, Ahn, Kim,
  Wilson, and John}}]{graphite-AA}
\bibinfo{author}{\bibfnamefont{J.-K.} \bibnamefont{Lee}},
  \bibinfo{author}{\bibfnamefont{S.-C.} \bibnamefont{Lee}},
  \bibinfo{author}{\bibfnamefont{J.-P.} \bibnamefont{Ahn}},
  \bibinfo{author}{\bibfnamefont{S.-C.} \bibnamefont{Kim}},
  \bibinfo{author}{\bibfnamefont{J.~I.} \bibnamefont{Wilson}},
  \bibnamefont{and} \bibinfo{author}{\bibfnamefont{P.}~\bibnamefont{John}},
  \bibinfo{journal}{The Journal of chemical physics}
  \textbf{\bibinfo{volume}{129}}, \bibinfo{pages}{234709}
  (\bibinfo{year}{2008}).

\bibitem[{\citenamefont{Uchida et~al.}(2014)\citenamefont{Uchida, Furuya,
  Iwata, and Oshiyama}}]{uchida-tbg-prb14}
\bibinfo{author}{\bibfnamefont{K.}~\bibnamefont{Uchida}},
  \bibinfo{author}{\bibfnamefont{S.}~\bibnamefont{Furuya}},
  \bibinfo{author}{\bibfnamefont{J.-I.} \bibnamefont{Iwata}}, \bibnamefont{and}
  \bibinfo{author}{\bibfnamefont{A.}~\bibnamefont{Oshiyama}},
  \bibinfo{journal}{Phys. Rev. B} \textbf{\bibinfo{volume}{90}},
  \bibinfo{pages}{155451} (\bibinfo{year}{2014}).

\end{thebibliography}

\end{document}